\documentclass[aps,pra,superscriptaddress]{revtex4-2}
\usepackage{epsfig,amssymb, amsmath,fancyhdr,upgreek,color,hyperref}
\newcommand{\bzeta}{{\boldsymbol \zeta}}
\newcommand{\bhata}{{\hat{\boldsymbol a}}}
\newcommand{\balpha}{{\boldsymbol \alpha}}

\begin{document}

\title{Entanglement distribution:  To herald or not to herald}

\author{Jeffrey H. Shapiro}
\email{jhs@mit.edu}
\affiliation{Department of Electrical Engineering and Computer Science, and Research Laboratory of Electronics, Massachusetts Institute of Technology, Cambridge, Massachusetts 02139 USA}
\author{Clark Embleton}
\affiliation{Department of Physics, Oregon Center for Optical, Molecular, and Quantum Science, University of Oregon, Eugene, Oregon 97403 USA}
\author{J. Gabriel Richardson}\affiliation{Department of Electrical and Computer Engineering, University of Maryland, College Park, MD 20742, USA}

\begin{abstract} 
High-rate, high-fidelity entanglement distribution is essential for the creation of a quantum internet, and spontaneous parametric downconverters (SPDCs) are, at present, the preferred sources of entangled signal-idler photon pairs for transmission to Alice and Bob's quantum nodes.  SPDCs using phase-matched spectral islands are especially attractive, in this regard, because they provide wavelength-division multiplexed signal-idler pairs with single-mode temporal behavior.  This paper compares the entanglement distribution rates of three islands-based systems.  Two use idler detections for heralding:  islands-based zero-added-loss multiplexing (ZALM), and an islands-based Sagnac SPDC source with signal-path erasure.  The third employs an unheralded Sagnac SPDC source.  For 90\% or lower heralding efficiencies,  ZALM's per-pump-pulse entanglement distribution rate exceeds that of the signal-path erasure source, and  both rates are inferior to unheralded operation's when all three systems employ $N_I$ spectral islands and allocate $N_M = N_I$ quantum memories to each pump pulse.   
These behaviors, however, must be weighed against the three systems' differing equipment requirements, e.g., ZALM requires a pair of perfectly-matched Sagnac sources, which is a significant burden not incurred by the signal-path erasure approach, and both heralded systems will suffer, in comparison with unheralded operation, if they cannot realize high heralding efficiencies.

\end{abstract}

\maketitle

\section{Introduction}
Spontaneous parametric downconverters (SPDCs) have long been workhorses for generating polarization entangled photon pairs~\cite{Kwiat1995,Kwiat1999,Kim2001,Kurtsiefer2001} and time-bin entangled photon pairs~\cite{Brendel1999,Marcikic2002,Marcikic2004}.   They have been used in loophole-free Bell-inequality tests of quantum mechanics~\cite{Giustina2015,Shalm2015}, and in record-setting demonstrations of long-distance entanglement distribution over optical fiber~\cite{Neumann2022} and from a low-earth-orbit satellite~\cite{Yin2017}.  Long-distance entanglement distribution is essential for there to be a quantum internet~\cite{Lloyd2004,Kimble2008,Wehner2018}, because distributed entanglement enables qubit teleportation~\cite{Bennett1993,Bouwmeester1997,Riebe2004}, which is the only  known means for error-free qubit transmission over channels with 3\,dB or greater loss~\cite{Weedbrook2012}.  In SPDC-based entanglement distribution the downconverter is typically pulse-pumped, to provide timing information, and operated at low brightness ($\sim$0.01 photon pair per pump pulse is the usual rule of thumb for unheralded operation), to avoid multipair events that can create erroneous quantum memory loads.  Taken together with the likely 10--20\,dB propagation losses of inter-node photon propagation, SPDC-based entanglement distribution will be woefully inadequate to meet the high-rate needs of a quantum internet \emph{unless} an efficient means for multiplexing many such sources can be developed.  
 
Prior SPDC multiplexing suggestions that relied on path switching at the source, e.g.\@ Refs.~\cite{Mower2011,Dhara2022}, suffer significant performance penalties from their switch losses.  A much better multiplexing prospect was proposed by Chen~\emph{et al}.~\cite{Chen2023}, viz., zero-added-loss multiplexing (ZALM).  ZALM uses wavelength-channelized partial Bell-state measurements (BSMs)~\cite{Braunstein1995} on the idler photons from a pair of Sagnac SPDCs'~\cite{Wong2006} polarization-entangled photon pairs.  The heralded signal photons are then sent to Alice and Bob's quantum receivers (QRXs) along with a classical signal identifying their wavelength~\cite{footnote1}.  Quantum-state wavelength conversion~\cite{Kumar1990,Huang1992} matches incoming light's wavelength at Alice and Bob's QRXs to that of their quantum memories, thus avoiding switch losses in ZALM's quantum transmitter (QTX) and justifying its ``zero-added-loss'' name.  Chen~\emph{et al}.\@ estimated that with 800 channels ZALM could realize quasideterministic ($\ge$25\% per pump burst~\cite{footnote2} probability), high-fidelity ($\ge$99\%) heralded generation of polarization-entangled biphotons.  ZALM, as proposed however, comes with a high technological burden: its partial BSM for 800 channels requires 3200 superconducting nanowire single-photon detectors (SNSPDs) with partial photon-number resolution (partial-PNR) capability, which distinguishes between 0, 1, or $>$1 detected photons.   

In a subsequent study~\cite{Shapiro2024}, Shapiro~\emph{et al}.\@ took a deep dive into ZALM theory, assuming conventional pulse pumping and channelization by means of dense wavelength-division multiplexing (DWDM) filters in the QTX's partial-BSM apparatus and in Alice and Bob's QRXs.  Their work exhibited a fundamental tradeoff between heralding efficiency and Bell-state fidelity, which Chen~\emph{et al}.'s scheme avoids by its pump burst's resulting in channels whose biphoton wave functions \emph{approximate} spectrally-factorable states.  So, inspired by Morrison~\emph{et al}.~\cite{Morrison2022}, whose domain-engineered $\chi^{(2)}$ crystal pumped by a transform-limited Gaussian pulse with
an appropriately chosen bandwidth realized a biphoton wave function with 8 discrete and spectrally factorable frequency bins (islands), Shapiro~\emph{et al}.~\cite{Shapiro2025} examined ZALM performance using Sagnac SPDCs with $N_I$ ideal phase-matched spectral islands.  Their results were remarkable in that, unlike Refs.~\cite{Chen2023,Shapiro2024}, the new analysis accounted for multipair events of all orders and losses in both the partial BSM and propagation to Alice and Bob's QRXs.  In addition, and more importantly, they showed that 41 spectral islands suffices to achieve 25\% per pump pulse heralded generation of polarization-entangled biphotons when the partial-BSM's efficiency is 90\%.  Moreover, that setup delivers $\ge$99.98\% Bell-state fraction with 99\% Bell-state fidelity to Alice and Bob's QRXs when there is 20\,dB propagation loss on each QTX-to-QRX connection.  Essential to Ref.~\cite{Shapiro2025}'s reducing Chen~\emph{et al}.'s 800 wavelength channels to 41 phase-matched spectral islands is its use of same-plus-cross-island (SPCI) heralding, i.e., declaring a herald when the partial BSM has idler clicks on appropriate detectors~\cite{footnote3}, \emph{regardless} of their wavelengths.  In contrast, Refs.~\cite{Chen2023,Shapiro2024} assume same-channel heralding, viz.,  both idler detections must occur in the same DWDM channel.   For both same-channel and SPCI heralding the signal photons' $H$-polarized and $V$-polarized components can be frequency converted to match what is needed for the intended quantum memories.  But, because the $H$-polarized and $V$-polarized components' frequency conversions must be done in a phase-locked manner, achieving that is more complicated---and considerably more challenging when the islands' collective bandwidth spans more than few THz~\cite{footnoteA}---for SPCI heralding than for same-island heralding.

In the present work we extend Shapiro~\emph{et al}.'s study~\cite{Shapiro2025} of ZALM using phase-matched spectral islands in several directions.  First, we compare Ref.~\cite{Shapiro2025}'s results with the performance achieved when its dual-Sagnac ZALM configuration is replaced by Chahine~\emph{et al}.'s simpler, single-Sagnac, signal-path erasure configuration ~\cite{Chahine2025} that also uses phase-matched spectral islands.  Because a simpler source may lead to higher heralding efficiency, this comparison is important, cf.\@ the performance of the 90\% efficient, 41-island source cited above being matched by a 95\% efficient, 20-island source, while an 80\%-efficient source requires 91 islands to meet those performance metrics~\cite{Shapiro2025}.  

Second, we compare the foregoing heralded approaches to unheralded operation with a single Sagnac source. We first make the heralded versus unheralded comparison  when at most a single herald is transmitted to Alice and Bob for a particular pump pulse.  In this case Alice and Bob's QRXs need only devote a single quantum memory to each pump pulse, and the unheralded source need only have a single phase-matched spectral island.   However,  because high heralding efficiency can enable a multi-island source's providing multiple heralds from a single pump pulse, we revisit heralded versus unheralded operation when Alice and Bob's QRXs commit $N_M$ memories to each pump pulse, and the unheralded source has $N_M$ phase-matched spectral islands.  Here we find that unheralded operation's per-pump-pulse entanglement distribution rates exceeds that of the heralded sources when $N_I=N_M$ and their heralding efficiencies are an optimistic 90\%.

Finally, we evaluate the impacts of dark counts in the heralded sources' single-photon detectors, and background light in free-space propagation from the QTX to Alice and Bob's QRXs.  
Chen~\emph{et al}.~\cite{Chen2023} addressed these issues, but the analyses therein need updating to apply to operation with phase-matched spectral islands.  

The remainder of this paper is organized as follows.  In Sec.~\ref{configs} we describe the configurations for: (1) the ZALM QTX with Ref.~\cite{Shapiro2025}'s dual-Sagnac source; (2) Chahine~\emph{et al}'s single-Sagnac QTX with idler-detection heralding and signal-path erasure; and (3) an unheralded single-Sagnac QTX.  There we also contrast the mode-conversion requirements for heralded and unheralded operation, and introduce Ref.~\cite{Shapiro2025}'s performance metrics, which will be used in what follows.  Section~\ref{performance} is the paper's heart, in which we evaluate and compare our three configurations' performance metrics, first when only a single herald is allowed for each pump pulse, and then when multiple heralds are permitted.  In Sec.~\ref{nonidealities} we turn to nonidealities that were not included in Ref.~\cite{Shapiro2025}, specifically, heralding-detector dark counts, and background light superimposed on the QTX light reaching Alice and Bob's QRXs.  Section~\ref{Discussion} wraps up the paper with some discussion and suggestions for follow-on work.  Derivations for some of the results presented in Secs.~\ref{performanceB}, \ref{performanceC}, and \ref{Background} are relegated to Appendices~\ref{AppendA}, \ref{AppendB}, and \ref{AppendC}, respectively.

\section{Configurations for Heralded and Unheralded Operation \label{configs}}

\begin{figure}[hbt]
    \centering
\includegraphics[width=4.5in]{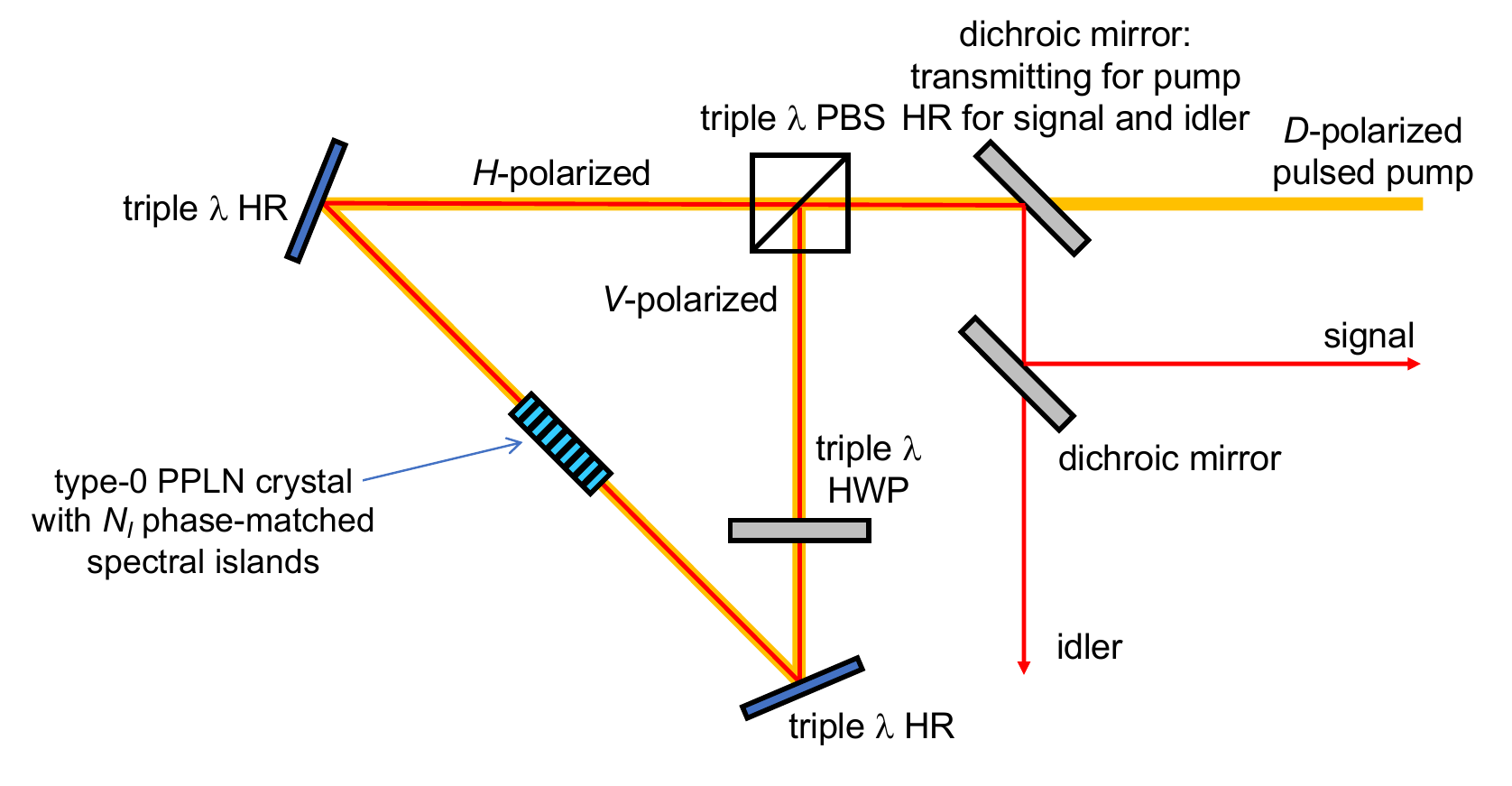}
\caption{Schematic of a Sagnac-configured SPDC source~\cite{Wong2006} of signal-idler biphotons suitable for use with $N_I\ge 1$ in islands-based heralded operation using either the dual-Sagnac source or the Chahine~\emph{et al}.\@ source.  A periodically-poled lithium niobate (PPLN) crystal with $N_I$ phase-matched spectral islands~\cite{footnote4} is bidirectionally pulse-pumped for type-0 nondegenerate phase matching.  $D$, $H$, and $V$: diagonal, horizontal, and vertical polarizations.  HR:  high reflector. $\lambda$:  wavelength.  PBS:  polarizing beam splitter.  HWP:  half-wave plate. \label{SagnacSource_fig}}    
\end{figure}

\begin{figure}[hbt]
    \centering
\includegraphics[width=1.5in]{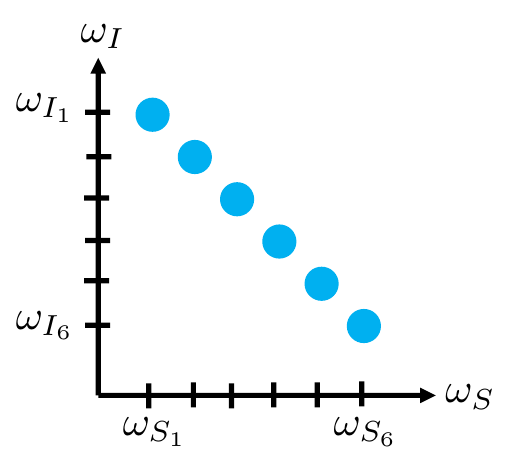}
   \caption{Sketch of the frequency-domain wave function for a biphoton produced by 6 identical, spectrally-factorable phase-matched spectral islands, with signal-idler center frequencies $\{(\omega_{S_n},\omega_{I_n}): n=1,2,\ldots,6\}$.
 \label{islands_sketch} }
\end{figure}

The bedrock for all of what follows is Fig.~\ref{SagnacSource_fig}'s islands-based SPDC source~\cite{Shapiro2024}.  It consists of a periodically-poled lithium niobate (PPLN) crystal with $N_I \ge 1$ phase-matched spectral islands. These islands are type-0 phase matched, and pulse-pumped in phase to produce signal-idler pairs that are in identical temporal modes except for their island-dependent center frequencies.  To achieve such a state, the spectral islands must be identical, nonoverlapping, and spectrally factorable, i.e., expressible as a product of functions of the signal frequency and the idler frequency, as sketched in Fig.~\ref{islands_sketch} for 6 islands.  

Restricting our attention to the foregoing temporal modes, and assuming operation in the SPDC's usual no-pump-depletion regime, the output state from the Sagnac source shown in Fig.~\ref{SagnacSource_fig} resulting from a single pump pulse is the following tensor product of two-mode squeezed-vacuum (TMSV) states,
\begin{equation}
|\psi\rangle_{SI} = \bigotimes_{n=1}^{N_I}\sum_{m=0}^\infty\sqrt{P_m}\,|m\rangle_{S_{n_H}}|m\rangle_{I_{n_H}} \bigotimes_{n'= 1}^{N_I}\sum_{m'=0}^\infty\sqrt{P_{m^\prime}}\,|m'\rangle_{S_{n'_V}}|m'\rangle_{I_{n'_V}},
\label{islands_state}
\end{equation}
where, as in Ref.~\cite{Shapiro2025}, source losses have been taken to be symmetrically distributed and ascribed either to the heralding detectors or QTX-to-QRX propagation, depending on the QTX configuration.
In Eq.~(\ref{islands_state}): $|m\rangle_{K_{n_P}}$ is the $m$-photon Fock state of the $P$-polarized ($P=H,V$) signal ($K=S$) and idler ($K=I$) modes emitted by the $n$th phase-matching island; and 
\begin{equation}
P_m \equiv \frac{(G-1)^m}{G^{m+1}},\mbox{ for $m=0,1,\ldots, \infty$},
\label{BoseEinstein}
\end{equation} 
is the Bose-Einstein probability distribution with $G$ being the squeezing gain, so that $G-1 > 0$ is the average number of signal-idler pairs emitted per SPDC-island per pump pulse.  

\subsection{Heralding Configuration for Islands-Based ZALM \label{configurationA}}
\begin{figure}[hbt]
    \centering
\includegraphics[width=4in]{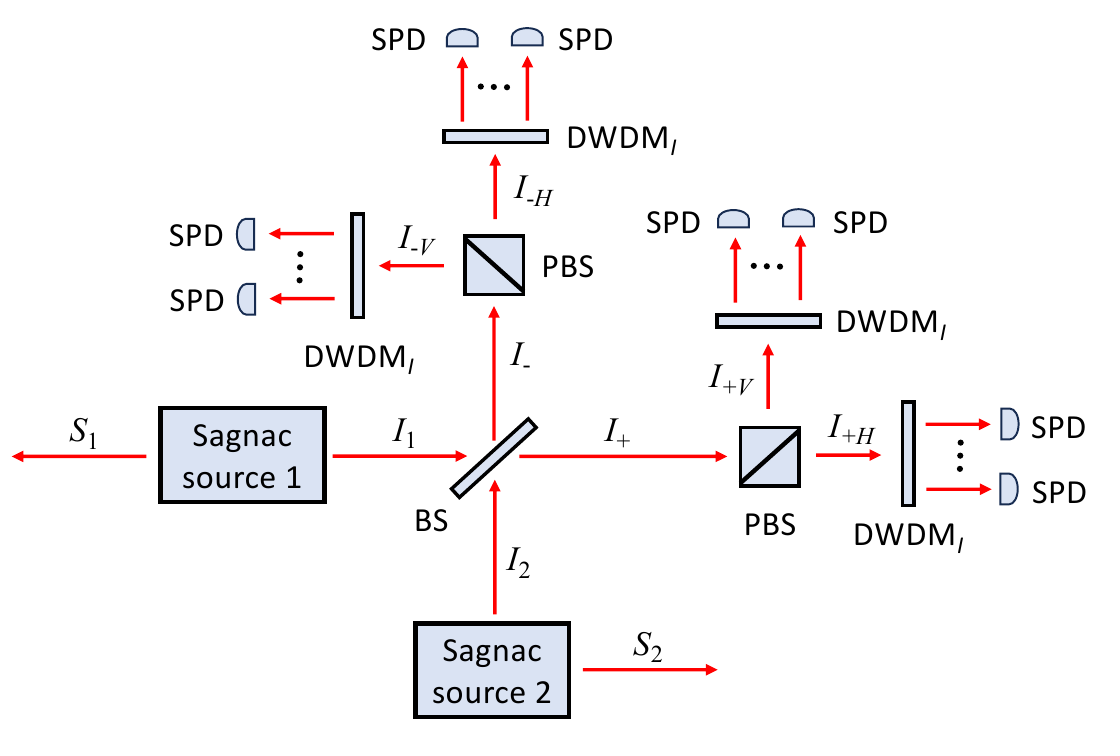}
    \caption{Schematic of islands-based ZALM's partial Bell-state measurement for heralding polarization-entangled photon pairs. Here $S_k$ and $I_k$ for $k = 1,2$ denote the signal ($S$) and idler ($I$) beams from the $k$th Sagnac source.  $I_\pm$ denote the idler-beam outputs from the 50--50 beam splitter (BS); $I_{\pm P}$ for $P= H, V$ denote the horizontally ($H$) and vertically ($V$) polarized outputs from the polarizing beam splitter (PBS) illuminated by $I_\pm$;  DWDM$_I$:  idler-beam dense wavelength-division multiplexing filter.  SPD:  single-photon detector.
 \label{partial-BSM_fig} }
\end{figure}

The heralding setup for ZALM using the dual-Sagnac source is shown in Fig.~\ref{partial-BSM_fig}.   Its two Sagnac sources are coherently pumped so that their joint state, from a single pump pulse, is $|\psi\rangle_{\bf SI} = |\psi\rangle_{S_1I_1} |\psi\rangle_{S_2I_2}$,  
where the individual Sagnacs' states are given by Eq.~(\ref{islands_state}).  Their idler beams are combined on a 50--50 beam splitter, polarization analyzed into their horizontally ($H$) polarized and vertically ($V$) polarized components by polarizing beam splitters (PBSs), and channelized by DWDM filters that direct the light from different islands to different single-photon detectors (SPDs).  The SPDs are assumed to have partial photon-number resolution (partial-PNR) capability, so that they can distinguish between 0, 1, or $>$1 detections, and to have no dark counts.  Their $\eta_T < 1$ quantum efficiencies combine all the losses on the idler paths.  A same-island herald for a single pump pulse is declared for the $n$th island when exactly two idler photons are detected from that island, with one being $H$ polarized and the other being $V$ polarized.  A cross-island herald for a single pump pulse is declared for islands $n$ and $m \ne n$ when an $H$-polarized photon is detected from the $n$th island and a $V$-polarized photon is detected from the $m$th island.  For $1\le n,m \le N_I$, if one detection occurs from Fig.~\ref{partial-BSM_fig}'s $I_+$ branch, and the other from its $I_-$ branch, then a $\psi^-_{nm}$ herald is declared,
where
\begin{equation}
|\psi^-\rangle_{{\bf S}_{nm}} = (|H\rangle_{S_{1_n}}|V\rangle_{S_{2_m}} - |V\rangle_{S_{1_m}}|H\rangle_{S_{2_n}})/\sqrt{2},
\label{psiminus}
\end{equation}
with $|P\rangle_{S_{k_j}}$ for $P= H,V$, $k=1,2$, and $j=n,m$ being the $P$-polarized single-photon state of the $S_{k_j}$ mode, is the polarization-singlet state.  Otherwise, i.e., if both idler detections occur on the same branch, a $\psi^+_{nm}$ herald is declared, where
\begin{equation}
|\psi^+\rangle_{{\bf S}_{nm}} = (|H\rangle_{S_{1_n}}|V\rangle_{S_{2_m}} + |V\rangle_{S_{1_m}}|H\rangle_{S_{2_n}})/\sqrt{2}.
\label{psiplus}
\end{equation}
Note that half of these heralds are \emph{false} heralds, in which signal photons are sent to either Alice or Bob but not to both~\cite{Shapiro2024}, hence the statements above read a ``herald is declared'', rather than the state is heralded.   

The performance analysis in Sec.~\ref{performanceA} employs same-plus-cross-island (SPCI) heralding, and assumes that Fig.~\ref{partial-BSM_fig}'s signal photons from Sagnacs 1 and 2 are transmitted to Alice and Bob's QRXs, respectively, through propagation channels with $\eta_R < 1$ transmissivities.  Also transmitted to them is a classical message,  which we assume is received without error, identifying the heralded state and the wavelengths of the $H$-polarized and $V$-polarized photons. 

\subsection{Heralding Configuration for Chahine~\emph{et al}.'s Islands-Based Source \label{configurationB}}
\begin{figure}[hbt]
    \centering
\includegraphics[width=3in]{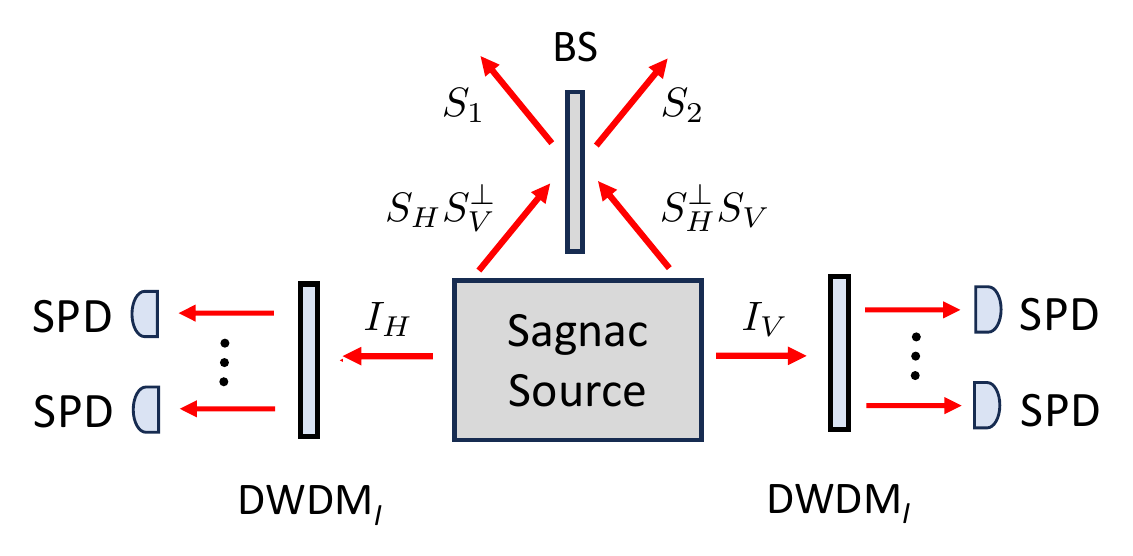}
    \caption{Schematic of Chahine~\emph{et al}.'s islands-based source of heralded polarization-entangled photon pairs.  $S^\perp_P$ for $P = V,H$ are auxiliary vacuum-state $P$-polarized modes needed to ensure that the $S_k = S_{k_H}S_{k_V}$ modes for $k=1,2$ have proper free-field commutators. DWDM$_I$: idler-beam dense wavelength-division multiplexing filter.  SPD:  single-photon detector.  BS: 50--50 beam splitter.
 \label{Chahine_source_fig} }
\end{figure}

The 50--50 beam splitter in Fig.~\ref{partial-BSM_fig} erases path information, so that every idler detection is equally likely to have come from either Sagnac, thus entangling the signal photons sent to Alice and Bob when one idler is detected from each source.  Chahine~\emph{et al}.'s heralded source, shown in Fig.~\ref{Chahine_source_fig}, uses a 50--50 beam splitter for the same path-erasure purpose, only now that occurs on the signals, rather than the idlers, and only one Sagnac source is required.  The islands-based Sagnac source from Fig.~\ref{SagnacSource_fig} produces $H$-polarized TMSV signal-idler states from its clockwise (cw) path through the PPLN crystal and $V$-polarized TMSV signal-idler states from its counterclockwise (ccw) path through that crystal. The $H$-polarized and $V$-polarized idlers from the cw and ccw sources' $N_I$ islands are channelized by a pair of DWDM filters and detected by SPDs with partial-PNR (0, 1, or $>$1) capabilities, no dark counts, and $\eta_T < 1$ quantum efficiencies.  At the same time, those sources' signal modes, along with auxiliary vacuum-state modes, are combined on a 50--50 beam splitter and then transmitted to Alice and Bob, respectively.  When cw source's $n$th-channel idler detector registers a single photon, and the ccw source's $m$th-channel idler detector registers a single photon, a $\psi^-_{nm}$ herald is declared.  The Chahine~\emph{et al}.\@ source is unable to herald  the other polarization-Bell states, and it too suffers from having half of its heralds be false heralds associated with signal photons being sent to Alice or to Bob but not to both.   As was the case for the ZALM source, there is also error-free classical transmission to Alice and Bob of the heralded wavelengths for the $H$-polarized and $V$-polarized photons, and our performance analysis for the Chahine~\emph{et al.}\@ source will assume SPCI heralding.

In short, the idler detections in the Chahine~\emph{et al}.\@ configuration identify which path (cw or ccw) they took through the PPLN crystal, but 50--50 combining of the signals from those paths erases the path information, hence the signals sent to Alice and Bob are polarization entangled if both contain photons.  In Sec.~\ref{performanceB} we will provide the details.  

\subsection{Configuration for Islands-Based Unheralded Operation \label{configurationC}}
\begin{figure}[hbt]
    \centering
\includegraphics[width=4.5in]{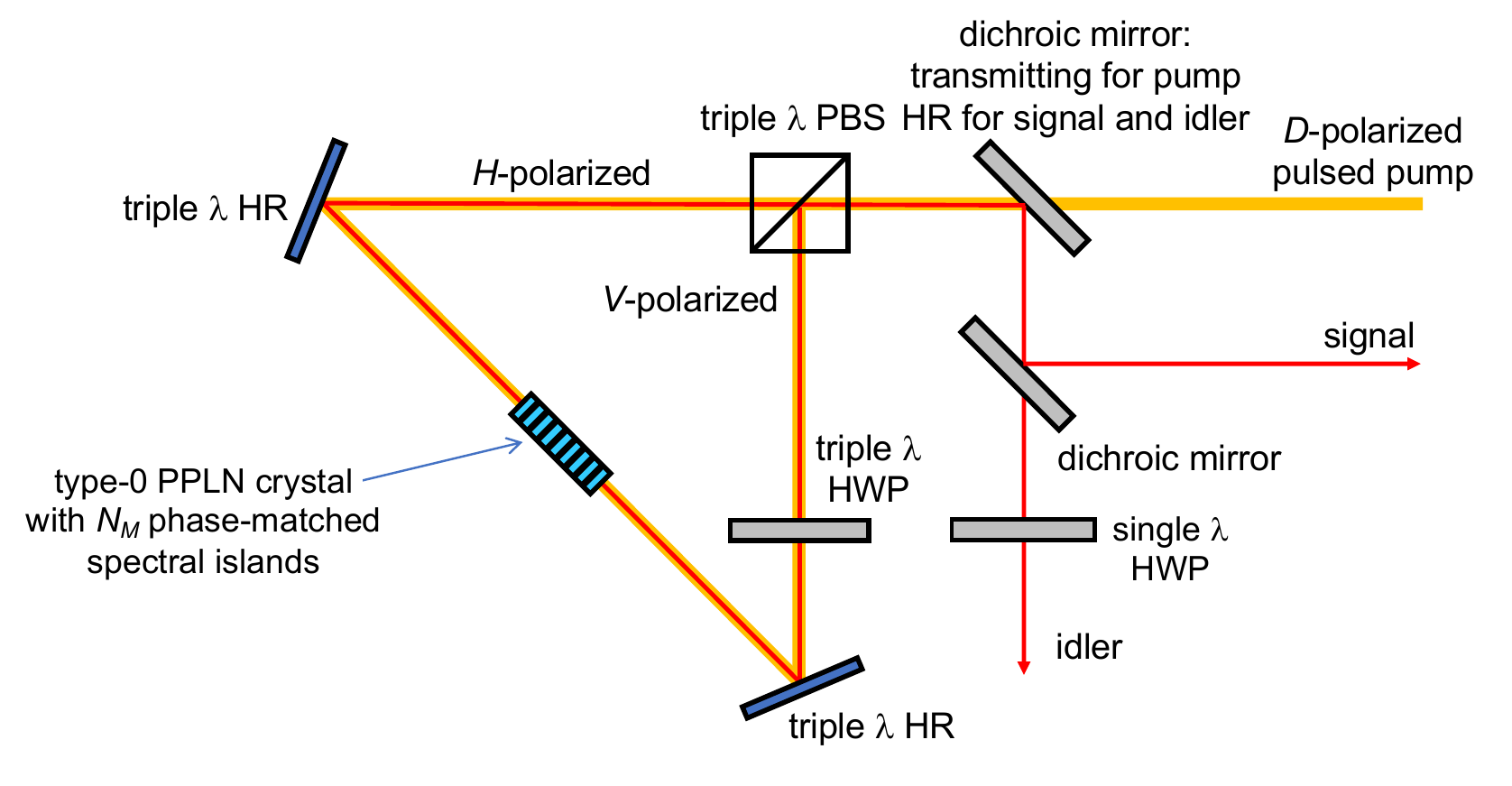}
    \caption{Schematic of islands-based source for unheralded distribution of polarization-entangled photon pairs. A periodically-poled lithium niobate (PPLN) crystal with $N_M$ phase-matched spectral islands~\cite{footnote5} is bidirectionally pulse-pumped for type-0 nondegenerate phase matching.  $D$, $H$, and $V$: diagonal, horizontal, and vertical polarizations.  HR:  high reflector. $\lambda$:  wavelength.  PBS:  polarizing beam splitter.  HWP:  half-wave plate. 
 \label{unheralded_fig} }
\end{figure}

The Sagnac source for islands-based unheralded operation is shown in Fig.~\ref{unheralded_fig}.  It differs from the Sagnac configuration used for the heralded approaches in two ways.  First, an idler-wavelength HWP has been added to the idler's output path, and second, the PPLN crystal now has $N_M$ phase-matched spectral islands, where $N_M \ge 1$ is the number of quantum memories Alice and Bob's QRXs assign to each pump pulse.  

As will be shown in Sec.~\ref{performanceC}, this source's signal-idler state resulting from a single pump pulse is 
\begin{equation}
|\psi\rangle_{SI} = \bigotimes_{n=1}^{N_M}\sum_{m=0}^\infty\sqrt{P_m}\,|m\rangle_{S_{n_H}}|m\rangle_{I_{n_V}} \bigotimes_{n'= 1}^{N_M}\sum_{m'=0}^\infty(-1)^{m^\prime}\sqrt{P_{m^\prime}}\,|m'\rangle_{S_{n'_V}}|m'\rangle_{I_{n'_H}},
\label{unheralded_islands_state}
\end{equation}
with source losses taken to be symmetrically distributed and ascribed to the QTX-to-QRX propagation paths.  The idler kets' $H\rightarrow V, V\rightarrow H$ polarization exchanges and the $(-1)^{m^\prime}$ factor are due to the HWP added to idler's output path.  We have added this HWP so that in the perturbative regime, when each island emits at most one biphoton, each of those biphotons is a polarization-singlet state, making it easy to apply Ref.~\cite{Shapiro2024}'s theory for loading a Duan-Kimble intra-cavity quantum memory~\cite{Duan2004}.  The unheralded source's signal and idler outputs from Fig.~\ref{unheralded_fig} are sent to Alice and Bob, respectively.

\subsection{Mode Conversion Requirements:  Heralded versus Unheralded Operation}
Although the main thrust of this paper is to compare three approaches to islands-based entanglement distribution to Alice and Bob's QRXs, with no consideration of loading their quantum memories, it still behooves us to give some high-level consideration to the different mode-conversion architectures they need to interface efficiently with those quantum memories.  Both Chen~\emph{et al}.~\cite{Chen2023} and Shapiro~\emph{et al}.~\cite{Shapiro2024} assumed same-island heralding with Alice and Bob each committing a single memory per pump pulse to the light they receive.  Thus tunable polarization-independent wavelength conversion from $\lambda_{S_n}$, the signal wavelength of the heralded channel, to $\lambda_M$, the wavelength of the quantum memory followed by temporal-mode shaping, to maximize coupling to the quantum memory, sufficed for their mode converters.  These requirements are easily stated, but quite demanding.  They need to be quantum-state preserving, i.e., near lossless with minimum noise injection.  Shapiro~\emph{et al}.~\cite{Shapiro2025} introduced SPCI heralding, while continuing to assume single-memory operation, and pointed out that Alice and Bob would now need to do polarization-dependent wavelength conversion with the same stringent requirements as same-island heralding.  For a single-island unheralded source the mode-conversion architecture is a simplified version of that assumed in Refs.~\cite{Chen2023,Shapiro2024}, viz., Alice's wavelength converter need only transform $\lambda_S$ to $\lambda_M$, while Bob's converts $\lambda_I$ to $\lambda_M$, where $\lambda_I\neq\lambda_S$ because of the PPLN SPDCs' nondegenerate phase matching.   

The present paper allows Alice and Bob to commit $N_M$ memories per pump pulse, where $1 < N_M \le N_I$.  It then allows the heralded sources to send more than one herald per pump pulse up to a maximum of $N_M$, and it assumes the unheralded source has $N_M$ phase-matched spectral islands.  A notional architecture of an $nm$th-island mode conversion that works for both islands-based ZALM and the Chahine~\emph{et al}.\@ source is shown in Fig.~\ref{mode-conversion1_fig}.  As will be seen below, unless the partial BSM's quantum efficiency, $\eta_T$, approaches unit value, the factor by which multiple-herald operation increases the entanglement distribution rate is well below $N_M$ unless $N_I$ is \emph{enormously} greater than $N_M$.  Furthermore, to gain multiple memories' full benefit the heralded systems' QRXs will need complicated path switching to accommodate any combination of $n$ and $m$ with $1\le n,m\le N_I$ for that enormous $N_I$ value.  We postpone discussion of the ensuing implementation complications until Section~\ref{Discussion}.  
\begin{figure}[hbt]
    \centering
\includegraphics[width=3.5in]{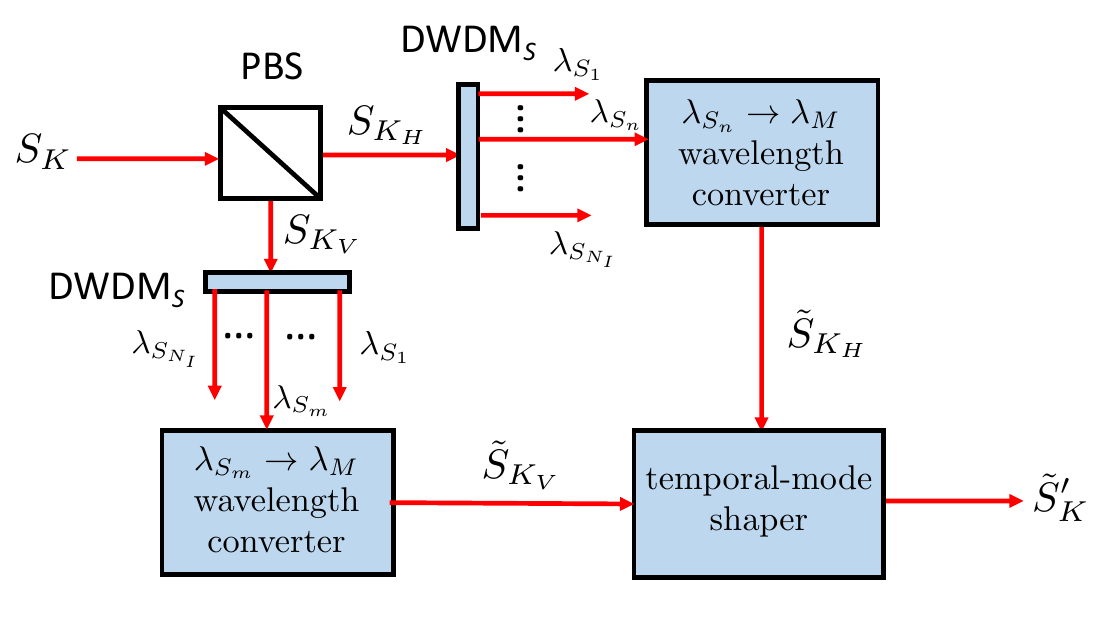}
    \caption{Notional architecture of an $nm$th-island mode conversion suitable for both  islands-based ZALM and the Chahine~\emph{et al}.\@ source.  $S_K$ for $K=A,B$ is the signal light arriving at Alice's ($A$) and Bob's ($B$) QRXs.  PBS:  polarizing beam splitter.  DWDM$_S$:  dense wavelength-division multiplexing filter for the $N_I$ signal wavelengths, $\{\lambda_{S_n}: n = 1,2,\ldots, N_I\}$. $\lambda_M$:  quantum memory wavelength.  $\tilde{S}'_K$ for $K = A,B$ is the mode-converted signal light at Alice and Bob's QRXs that is sent to those terminals' quantum memories.   Omitted from this figure is the path-switching network that routes the heralded-channel outputs from the DWDM$_S$ filters to mode-conversion modules.  \label{mode-conversion1_fig} }
\end{figure}
 
\begin{figure}[hbt]
    \centering
\includegraphics[width=4.25in]{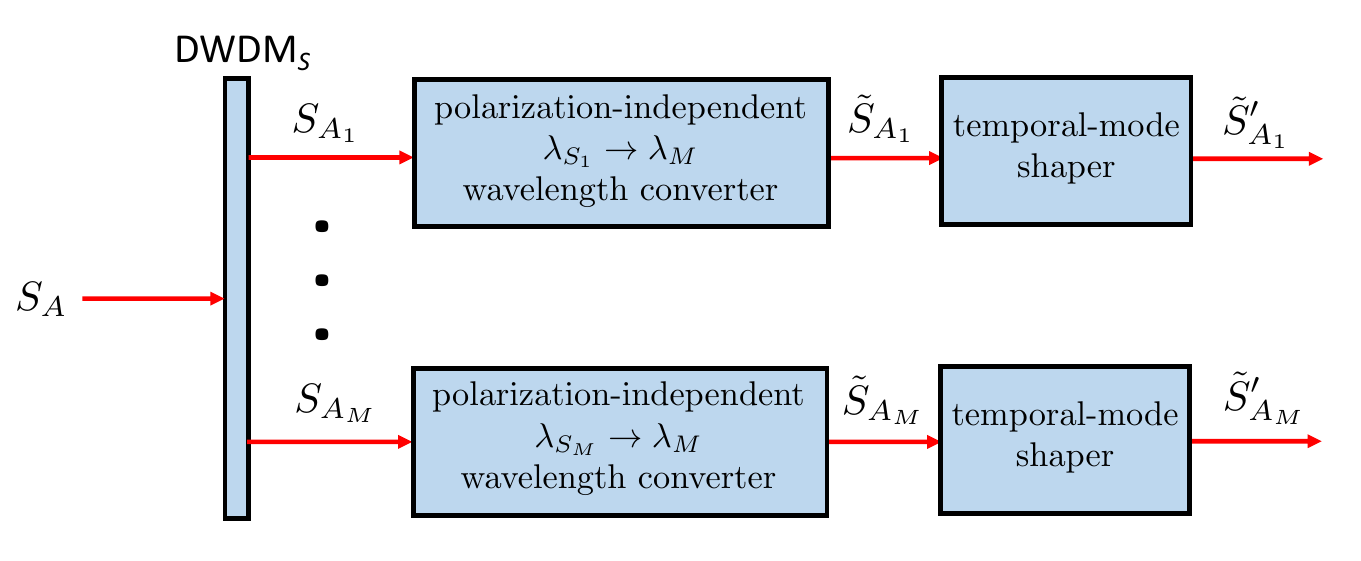}
    \caption{Wavelength-division multiplexed architecture of Alice's QRX for an $N_M$-island unheralded source.  $S_A$ is the signal light arriving at Alice's QRX.  DWDM$_S$:  dense wavelength-division multiplexing filter for the $N_M$ signal wavelengths, $\{\lambda_{S_n} : n = 1,2,\ldots, N_M\}$. $\lambda_M$:  quantum memory wavelength.  $\tilde{S}'_{A_n}$ is the $n$th-island mode-converted  signal light that is sent to Alice's $n$th-island quantum memory. 
 \label{mode-conversion2_fig} }
\end{figure}

In contrast to the mode-conversion complications encountered by our heralded sources when $N_I \gg N_M > 1$, and the modest benefit that accrues, at realistic $N_I$ values, the mode-conversion architecture for an unheralded islands-based source with $N_M >1$ islands is straightforward wavelength-division multiplexing, whose entanglement distribution rate scales linearly with $N_M$.  Figure~\ref{mode-conversion2_fig} shows the mode-conversion architecture for the signal light ($S_A$) arriving at Alice's QRX.  The same architecture applies to the idler light ($I_B$) arriving at Bob's QRX, with the signal-wavelength DWDM filter replaced by the idler-wavelength DWDM filter, and the $n$th-island wavelength conversion now being from $\lambda_{I_n}$ to $\lambda_M$.
 
\subsection{Figures of Merit}
The principal goal of this paper is to identify the most promising choice---among the three contenders under consideration---for maximizing the rate of high-quality entanglement distribution under given conditions.  In making that identification we need metrics for ``high quality'', and, for fairness, we need to be equitable in the resources we grant our contenders, e.g., the number of islands, the number quantum memories, and the complexity of the required mode converters.  Furthermore, harking back to Chen~\emph{et al}., we would like to realize quasideterministic entanglement generation for our heralded sources.  The present section draws from Ref.~\cite{Shapiro2024} for this paper's metrics and its definition of ``quasideterministic''.

We start with the notion of a \emph{loadable} $nm$th-island state, i.e., a joint state of the $n$th-island $H$-polarized light and the $m$th-island $V$-polarized light arriving at Alice and Bob's QRXs from a single pump pulse that delivers one or more photons to each QRX.  For our heralded sources we define $\Pr({\rm loadable}_{nm})$ to be the \emph{conditional} probability of a loadable state given that an $nm$th herald has occurred, where we allow both same-island ($n=m$) and cross-island ($n\ne m$) heralds, see Secs.~\ref{performanceA} and \ref{performanceB} for the details.  For our $N_M$-island unheralded source, $\Pr({\rm loadable}_n)$ is the \emph{unconditional} probability of an $n$th-island loadable state arriving at Alice and Bob's QRXs.  

Our primary performance metrics, the Bell-state fraction and the Bell-state fidelity, are defined as follows.  For the heralded sources, let $\Pr({\rm Bell}_{nm})$ be the conditional probability that the source delivers an $nm$th-island polarization-Bell state to Alice and Bob given that an $nm$th-island herald has occurred.  Similarly, for the $N_M$-island unheralded source, let $\Pr({\rm Bell}_n)$ be the unconditional probability that the source delivers an $n$th-island polarization-Bell state to Alice and Bob.   The Bell-state fractions for the heralded and unheralded
sources are then
\begin{equation}
\mathcal{B}_{nm} \equiv \frac{\Pr({\rm Bell}_{nm})}{\Pr({\rm loadable}_{nm})},
\end{equation}
and 
\begin{equation}
\mathcal{B}_n \equiv \frac{\Pr({\rm Bell}_n)}{\Pr({\rm loadable}_n)},
\end{equation}
i.e., the fraction of the loadable states that lie in the Hilbert space spanned by the polarization-Bell states for the modes in question.  We will see in Secs.~\ref{performanceA}--\ref{performanceC} that the loadable-state and Bell-state probabilities are independent of the island indices. Hence so too are their Bell-state fractions.  Thus in Sec.~\ref{performance}, where we focus on a single $nm$th-island state for the heralded sources and a single $n$th-island state for the unheralded source, we shall suppress mode indices in our analyses. 

For the ZALM source the $nm$th-island Bell-state fidelity is defined to be
\begin{equation}
\mathcal{F}^\pm_{nm} \equiv \frac{\Pr(|\psi^\pm\rangle_{{\bf S}_{nm}}{\mbox{ delivered} \mid \psi^\pm_{nm}\mbox{ herald})}}{\Pr({\rm Bell}_{nm})},
\end{equation}
and that for Chahine~\emph{et al}.'s source is defined to be
\begin{equation}
\mathcal{F}_{nm} \equiv \frac{\Pr(|\psi^-\rangle_{{\bf S}_{nm}}{\mbox{ delivered} \mid \psi^-_{nm}\mbox{ herald})}}{\Pr({\rm Bell}_{nm})},
\end{equation}
where ${\bf S}_{nm} \equiv S_{A_{n_H}}S_{A_{m_V}}S_{B_{n_H}}S_{B_{m_V}}$ are the relevant $nm$th-island signal modes at Alice's ($A$) and Bob's ($B$) QRXs.
For the $N_M$-island unheralded source the $n$th-island Bell-state fidelity is
\begin{equation}
\mathcal{F}_n \equiv \frac{\Pr(|\psi^-\rangle_{{\bf S}_{A_n}{\bf I}_{B_n}}{\mbox{ delivered})}}{\Pr({\rm Bell}_n)},
\end{equation}
where ${\bf S}_{A_n}{\bf I}_{B_n} \equiv S_{A_{n_H}}S_{A_{n_V}}I_{B_{n_H}}I_{B_{n_V}}$ are the relevant modes at Alice and Bob's QRXs.

As noted for the Bell-state fractions, the preceding Bell-state fidelities will turn out to be independent of the island indices and, for the ZALM source, whether the herald is $\psi^+_{nm}$ or $\psi^-_{nm}$.  So Sec.~\ref{performance} will suppress the fidelities' island indices.  We will also do so in the rest of this section for both the Bell-state fractions and the Bell-state fidelities.  

Before moving on to the per-pump-pulse entanglement distribution rate, it is worth commenting on the importance of the Bell-state fraction and the Bell-state fidelity.  When $1-\mathcal{B} \ll 1$ we can reasonably approximate the heralded sources' conditional density operators, $\hat{\rho}_{S_AS_B\mid \psi^-{\rm \,herald}}$, $\hat{\rho}_{S_AS_B\mid \psi^+{\,\rm herald}}$, by their normalized (trace = 1) projections onto the Hilbert space, $\mathcal{H}_B$, spanned by the polarization-Bell states, and likewise for the unconditional density operator, $\hat{\rho}_{S_AI_B}$ for unheralded operation.  In these cases Raymer~\emph{et al.}~\cite{Raymer2024} provides a rigorous derivation for the state-dependent reflectivities of an intra-cavity quantum memory, which we used in Ref.~\cite{Shapiro2024} to treat Duan-Kimble loading for the DWDM-channelized ZALM source in the perturbative regime.  Thus Ref.~\cite{Raymer2024} will enable us to do the same for islands-based ZALM, for Chahine~\emph{et al}.'s signal-path erasure scheme, and for the unheralded islands-based source.  To our knowledge no such rigorous treatment for those reflectivities exists for a significantly subunit $\mathcal{B}$.  A similar analysis benefit may also accrue for emissive quantum memories~\cite{Knall2022,Bersin2024} for islands-based entanglement distribution with near-unit Bell-state fraction.  Moreover, the Bell-state fidelity is clearly the figure of merit for the quality of the entanglement distributed to Alice and Bob's QRXs in this near-unit $\mathcal{B}$ regime. 

A heralded source's rate of distributed entanglement, in ebits per pump pulse, when Alice and Bob have $N_M>1$ memories available and the QTX can transmit up to $N_M$ heralds per pump pulse, is 
\begin{equation}
\mathcal{R}_E = \mathcal{R}_P\,\mathbb{E}[\mathcal{H}(N_M)]\,\mathcal{F}\Pr({\rm Bell}).
\label{ReMultiple}
\end{equation}
In this expression $\mathcal{R}_P$ is the pump's pulse-repetition rate, and 
\begin{equation}
\mathbb{E}[\mathcal{H}(N_M)] = \sum_{k=0}^{N_M-1}k\,p_k + N_M\sum_{k=N_M}^{N_I}p_k,
\mbox{ where $N_M \le N_I$},
\label{E(H)}
\end{equation}
 is the average number of SPCI heralds per pump pulse, with  $p_k$ being the probability that a single pump pulse results in $k$ SPCI heralds from the source's $N_I \ge N_M$ spectral islands~\cite{footnote6}.  Shapiro~\emph{et al}.~\cite{Shapiro2025} only treated single-memory operation, for which 
\begin{equation}
\mathcal{R}_E = \mathcal{R}_P\,(1-p_0)\,\mathcal{F}\Pr({\rm Bell}),
\label{ReSingle}
\end{equation}
where
\begin{equation}
p_0 = 2\!\left[1-\sqrt{\Pr(\mathcal{H}_{nm})}\right]^{N_I} - \left[1-\sqrt{\Pr(\mathcal{H}_{nm})}\right]^{2N_I},
\label{p0defn}
\end{equation}
with $\Pr(\mathcal{H}_{nm})$, the per-pump-pulse probability of an $nm$th-island herald, being independent of the island indices.  Expressions for $\Pr(\mathcal{H}_{nm})$ and $p_k$ will be given in Sec.~\ref{performanceA} for islands-based ZALM and in Sec.~\ref{performanceB} for the Chahine~\emph{et al}.\@ source.  

For unheralded islands-based operation the per-pump-pulse entanglement distribution rate takes the simpler form
\begin{equation}
\mathcal{R}_E = N_M\mathcal{R}_P\,\mathcal{F}\Pr({\rm Bell}),
\end{equation}
whose linear growth with increasing $N_M$ is due to this approach's being straightforward wavelength-division multiplexing.    

We close this section with our definition of quasideterministic entanglement generation, which we will only apply to single-memory operation.  When the heralded sources may only send one herald per pump pulse, the per-pump-pulse probability of their generating the heralded entangled state is
\begin{equation}
\Pr({\rm ent}) = \left.\mathcal{R}_E\right|_{\eta_R = 1}/\mathcal{R}_P \le 1/2,
\label{quasideterministic}
\end{equation} 
where the upper bound follows because $(1-p_0)\mathcal{F}\le 1$ by definition, and $\left.\Pr({\rm Bell})\right|_{\eta_R = 1}\le 1/2$ as half the heralds are false heralds.   In fact, the upper bound is tight, because $\eta_T =1$ gives $\mathcal{F} =1$ and $\left.\Pr({\rm Bell})\right|_{\eta_R = 1} = 1/2$, and $N_I\rightarrow \infty$ gives $p_0\rightarrow 0$.  For this reason we will take $\Pr({\rm ent}) = 1/4$ as islands-based ZALM's threshold for quasideterministic entanglement generation~\cite{footnote7}.  We will adhere to this definition, and also apply it to Chahine~\emph{et al}.'s signal-path erasure source, for single-herald operation.  
Single memory operation with the islands-based unheralded source will not provide quasideterministic performance because unheralded sources restrict their per-pump-pulse probability of emitting a biphoton to $\sim$0.01 to suppress multipair events.  

\section{Performance Analyses and Comparisons \label{performance}}
This section provides explicit results for $\Pr({\rm loadable})$, $\Pr({\rm Bell})$, $\mathcal{B}$, and $\mathcal{F}$ for all three configurations under consideration, as well as explicit results for $\Pr(\mathcal{H}_{nm})$ and $p_k$ for the heralded configurations.  In addition, for the heralded sources, it presents normalized (trace = 1) projections of $\hat{\rho}_{S_AS_B\mid \psi^-{\rm \,herald}}$ and $\hat{\rho}_{S_AS_B\mid\psi^+{\rm herald}}$ onto the Bell-state Hilbert space $\mathcal{H}_B$, and the corresponding normalized projection of $\hat{\rho}_{S_AI_B}$ for the unheralded source.

With those quantities in hand this section goes on to compare the relative merits of our contenders for high-quality, high-rate entanglement distribution, focusing on their per-pump-pulse distribution rates.  Before proceeding, however, it is worth remembering that our heralded configurations will both employ SPCI heralding.  That said, the results given for $\mathcal{B}$ and $\mathcal{F}$ being derived for an $nm$th-island herald, but \emph{not} depending on the $n$ and $m$ values, implies that they apply unchanged to same-island heralding.  The other heralded-configuration results presented in this section \emph{do} require SPCI heralding.
\subsection{Islands-Based ZALM Operation \label{performanceA}}
Virtually all the results we need for islands-based ZALM operation are available in Shapiro~\emph{et al.}~\cite{Shapiro2025}, with the exception being $p_k$, the per-pump-pulse probability of their being $k$ heralds when $1< N_M \le N_I$.  Thus, this section first introduces the characteristic function approach employed in Ref.~\cite{Shapiro2025}, which will be used in Secs.~\ref{performanceB} and \ref{performanceC}, then reprises Ref.~\cite{Shapiro2025}'s results, and finally derives the multi-memory quantity $p_k$.

The route employed in Ref.~\cite{Shapiro2025} to find the Bell-state fraction and Bell-state fidelity for the islands-based ZALM QTX started from the anti-normally-ordered characteristic function associated with $\hat{\rho}_{\bf SI}$, the density operator for the signal, ${\bf S} = S_1S_2$, and idler, ${\bf I} = I_1I_2$, modes of a single island, 
\begin{equation}
\chi_A^{\rho_{\bf SI}}(\bzeta) \equiv 
{\rm Tr}\!\left[\hat{\rho}_{\bf SI}e^{-\bzeta^\dagger\bhata}e^{\bhata^\dagger\bzeta}\right],
\label{antinorm_start}
\end{equation} 
where $\bzeta^\dagger \equiv \left[\begin{array}{cc}\bzeta_S^\dagger & \bzeta_I^\dagger\end{array}\right]$ and $\bhata^\dagger \equiv \left[\begin{array}{cc}\bhata_S^\dagger & \bhata_I^\dagger\end{array}\right]$ with 
\begin{equation}
\bzeta^\dagger_K \equiv \left[\begin{array}{cccc}\zeta^*_{K_{1_H}} & \zeta^*_{K_{1_V}} & \zeta^*_{K_{2_H}} & \zeta^*_{K_{2_V}}\end{array}\right],\mbox{ for $K=S,I$}, 
\end{equation}
and
\begin{equation}
\bhata_K^\dagger \equiv \left[\begin{array}{cccc}\hat{a}^\dagger_{K_{1_H}} & \hat{a}^\dagger_{K_{1_V}} & \hat{a}^\dagger_{K_{2_H}} & \hat{a}^\dagger_{K_{2_V}} \end{array}\right], \mbox{ for $K=S,I$,}
\end{equation}
being four-dimensional (4D) row vectors of complex-valued parameters and photon creation operators, respectively.  The density operator $\hat{\rho}_{\bf SI}$ can be recovered from its anti-normally-ordered characteristic function via the 16D operator-valued inverse Fourier transform~\cite{footnote8},
\begin{equation}
\hat{\rho}_{\bf SI} = \int\!\frac{{\rm d}^{16}\bzeta}{\pi^8}\,\chi_A^{\rho_{\bf SI}}(\bzeta)e^{-\bhata^\dagger\bzeta}e^{\bzeta^\dagger\bhata}.
\label{inverse_transform}
\end{equation}  The value of the transform-domain approach is that the anti-normally ordered characteristic functions for linear transformations of $\bhata$ and $\bhata^\dagger$ are trivial to find.  Thus accounting for Fig.~\ref{partial-BSM_fig}'s 50--50 beam splitter and its polarization beam splitters, as well as the beam-splitter models for its SPDs' subunit quantum efficiencies, is easily accomplished. Projecting the operator-valued inverse Fourier transform of the characteristic function so obtained onto the photon states for the heralding events then yields the unnormalized conditional density operator of the signal light sent to Alice and Bob given the heralding event that occurred.  Loss en route to Alice and Bob's QRX is another easily accounted for beam-splitter transformation of the unnormalized anti-normally ordered characteristic function for the transmitted signals. Given a $\psi^-$ herald from $I'_{\pm H}I'_{\mp V}$ detections, where the $I'$ modes account for the subunit quantum efficiencies of Fig.~\ref{partial-BSM_fig}'s SPDs, the anti-normally ordered characteristic function for the unnormalized density operator of Alice and Bob's received signal modes, $\tilde{\bf S} = S_AS_B$, is~\cite{Shapiro2025}
\begin{align}
&\chi_A^{\tilde{\rho}_{\tilde{\bf S}\mid I'_{\pm H}I'_{\mp V}}}(\bzeta) 
= \frac{[\eta_T(G-1)]^2}{[\eta_T(G-1)+1]^6}\nonumber \\[.05in]
&\,\,\times e^{-\bzeta^\dagger \bzeta/N'_S}\left[1 - \frac{\eta_R|\zeta_{A_H}\pm\zeta_{B_H}|^2}{2N_S}\right] 
 \left[1 - \frac{\eta_R|\zeta_{A_V}\mp\zeta_{B_V}|^2}{2N_S}\right] 
= \Pr(I'_{\pm H}I'_{\mp V})\,\chi_A^{\rho_{\tilde{\bf S}\mid I'_{\pm H}I'_{\mp V}}}(\bzeta),
\label{prebackgroundZALM}
 \end{align}  
with $N_S \equiv [\eta_T(G-1)+1]/G$, and $N_S' \equiv [\eta_R/N_S + (1-\eta_R)]^{-1}$. 
Projections of the unnormalized characteristic function's operator-valued inverse Fourier transform onto the loadable and Bell states then lead to the following results:
\begin{equation}
\Pr({\rm loadable}) = 1-2N_S'^2\left[1-\frac{\eta_RN_S'}{2N_S}\right]^2  + N_S'^4\!\left[1-\frac{\eta_RN_S'}{N_S}\right]^2
\label{loadableZALM},
\end{equation}
\begin{equation}
\Pr(|\psi^-\rangle_{\tilde{\bf S}}\mid \psi^-) = 
\frac{N_S'^4}{2}\!\left[2(1-N_S')^2 -\frac{2\eta_R(3N_S'^3-5N_S'^2+2N_S')}{N_S} + \frac{\eta_R^2(5N_S'^4-6N_S'^3+2N_S'^2)}{N_S^2}\right],
\label{PrCorrectZALM}
\end{equation}
and
\begin{align}
\Pr&(|\psi^+\rangle_{\tilde{\bf S}}\mid \psi^-) = 
\Pr(|\phi^\pm\rangle_{\tilde{\bf S}}\mid \psi^-)
\nonumber \\[.05in]
& = \frac{N_S'^4}{2}\!\left[2(1-N_S')^2 -\frac{2\eta_R(3N_S'^3-5N_S'^2+2N_S')}{N_S} + \frac{\eta_R^2(4N_S'^4-6N_S'^3+2N_S'^2)}{N_S^2}\right].
\label{PrErrorZALM}
\end{align}
As shown in Ref.~\cite{Shapiro2025}, the right-hand sides of Eqs.~(\ref{PrCorrectZALM}) and (\ref{PrErrorZALM}) also apply to the $\psi^+$ heralds, from $I'_{\pm H}I'_{\pm V}$ detections, when their left-hand sides are replaced by $\Pr(|\psi^+\rangle_{\tilde{\bf S}}\mid \psi^+)$ and $\Pr(|\psi^-\rangle_{\tilde{\bf S}}\mid \psi^+) = 
\Pr(|\phi^\pm\rangle_{\tilde{\bf S}}\mid \psi^+)$, respectively.   

From the preceding equations we now have that
\begin{align}
\Pr&({\rm Bell})  = \Pr(|\psi^\pm\rangle_{\tilde{\bf S}}\mid \psi^\pm)  + 
\Pr(|\psi^\mp\rangle_{\tilde{\bf S}}\mid \psi^\pm) + 
\Pr(|\phi^\pm\rangle_{\tilde{\bf S}}\mid \psi^\pm) +  \Pr(|\phi^\mp\rangle_{\tilde{\bf S}}\mid \psi^\pm) \nonumber \\[.05in]
&= 2N_S'^4\left[2(1-N_S')^2 -\frac{2\eta_R(3N_S'^3-5N_S'^2+2N_S')}{N_S} + \frac{\eta_R^2(4N_S'^4-6N_S'^3+2N_S'^2)}{N_S^2}\right] + \frac{\eta_R^2N^{\prime 8}_S}{2N_S^2}, \label{BellZALM} 
\end{align}
$\mathcal{B} = \Pr({\rm Bell})/\Pr({\rm loadable})$, and $\mathcal{F} =  \Pr(|\psi^\pm\rangle_{\tilde{\bf S}}\mid \psi^\pm)/\Pr({\rm Bell})$.  

The special case of a lossless partial BSM is especially interesting.  In particular, when $\eta_T = 1$ we get $N_S = N_S' = 1$, implying that
\begin{equation}
\Pr({\rm loadable}) = \Pr({\rm Bell}) = \Pr(|\psi^\pm\rangle_{\tilde{\bf S}}\mid \psi^\pm) = \eta_R^2/2,
\end{equation}
and hence $\mathcal{B} = \mathcal{F} = 1$.  Setting $\eta_R = 1$ in this expression shows that islands-based ZALM with a lossless partial BSM saturates Eq.~(\ref{quasideterministic})'s upper bound on the per-pump-pulse entanglement generation probability for single-herald operation when $p_0\rightarrow 0$, and its per-pump-pulse distribution rate is only reduced from $(1-p_0)\mathcal{R}_P/2$ by the probability, $\eta_R^2$, that the heralded Bell-state's two photons reach Alice and Bob's QRXs.

Another important result from Ref.~\cite{Shapiro2025} is its demonstration that $\hat{\rho}^{(\mathcal{H}_B)}_{\tilde{\bf S}\mid \psi^\pm}$, the normalized projection of $\hat{\rho}_{\tilde{\bf S}\mid\psi^\pm}$ onto the Bell-state Hilbert space, is diagonal in the Bell basis, and so is easily written down from Eqs.~(\ref{PrCorrectZALM}) and (\ref{PrErrorZALM}).  
Equation~(\ref{PrErrorZALM}) then shows that $\hat{\rho}^{(\mathcal{H}_B)}_{\tilde{\bf S}\mid \psi^\pm}$ is a Werner state.

Turning our attention to the multi-herald case, we have that an $nm$th-island herald occurs when one $H$-polarized $n$th-island idler photon and one $V$-polarized $m$th-island idler photon are detected.  Because all the idler states are independent, identically distributed (iid) thermal states, whose $\eta_T<1$ photon counting statistics are Bose-Einstein with mean $\eta_T(G-1)$, it follows that~\cite{Shapiro2025}
\begin{equation}
\Pr(\mathcal{H}_{nm}) = \frac{4[\eta_T(G-1)]^2}{[\eta_T(G-1)+1]^6}.
\label{ZALMheralding}
\end{equation} 
This expression can be decomposed into $\Pr(\mathcal{H}_{H_n})\Pr(\mathcal{H}_{V_m})$,
where
\begin{equation}
\Pr(\mathcal{H}_{P_j}) \equiv \frac{2\eta_T(G-1)}{[\eta_T(G-1)+1]^3}, 
\end{equation}
gives the probabilities for detecting one $H$-polarized $n$th-island idler photon ($P_j = H_n$), and for detecting one $V$-polarized $m$th-island idler photon ($P_j = V_m$).  Equation~(\ref{ReSingle})'s $(1-p_0)$ factor is then easily understood as the probability of their being at least one $nm$th herald because
\begin{equation}
p_0 = \Pr(\mbox{no $H_n$ detection}) +\Pr(\mbox{no $V_m$ detection}) - \Pr(
\mbox{no $H_n$ detection \& no $V_m$ detection})
\label{simple_p0}
\end{equation}
is the probability of their being no $nm$th-island herald.  In order to make our multi-memory derivations immediately applicable to Chahine~\emph{et al}.'s source, the rest of this section's work will use $q$ to denote $\Pr(\mathcal{H}_{P_j})$ for $P_j = H_n, V_m$.  The answers we obtain here will apply to islands-based ZALM by setting $q= 2\eta(G-1)/[\eta(G-1)+1]^3$.  Section~\ref{performanceB} will show that a different value of $q$ will give the corresponding answers for Chahine~\emph{et al}.'s source.   The analysis proceeds as follows.

For a single pump pulse, let $z_P$, for $P= H,V$, denote the number of islands from which a single $P$-polarized photon has been detected. The number of $nm$th-island heralds from that pump pulse is then $z \equiv \min(z_H,z_V)$.  Because the $H$-polarized and $V$-polarized photon detections 
from all $N_I$ islands are iid Bose-Einstein random variables whose single-photon detection probability equals $q$, we have that $z_H$ and $z_V$ are iid binomial-distributed random variables with success probability $q$, i.e., their common probability mass function is
\begin{equation}
P_{z_P}(\ell) = \frac{N_I!}{\ell!(N_I-\ell)!}\,q^\ell(1-q)^{N_I-\ell},\mbox{ for $\ell = 0,1,\ldots,N_I$, and $P = H,V$}.
\end{equation}
From this result we get
\begin{align}
p_k &\equiv \Pr(z = k) = \Pr(z \ge k) - \Pr(z\ge k+1) \\[.05in]
&= \left[\sum_{\ell = k}^{N_I}P_{z_H}(\ell)\right]\!\left[\sum_{\ell'= k}^{N_I}P_{z_V}(\ell')\right] -
 \left[\sum_{\ell = k+1}^{N_I}P_{z_H}(\ell)\right]\!\left[\sum_{\ell'= k+1}^{N_I}P_{z_V}(\ell')\right] \\[.05in]
 & = \frac{N_I!}{k!(N_I-k)!}\,q^k(1-q)^{N_I-k}\!\left\{\frac{N_I!}{k!(N_I-k)!}\,q^k(1-q)^{N_I-k}\right. 
 \nonumber \\[.05in]  
 & \,\, +2\left.\left[\sum_{\ell = k+1}^{N_I}\frac{N_I!}{\ell!(N_I-\ell)!}\,q^\ell(1-q)^{N_I-\ell}\right]\right\}, \mbox{ for $k = 0,1,\ldots, N_I$.}
 \label{simple_pk}
 \end{align}
Setting $k=0$ in the above result, and doing a little double bookkeeping inside the bracket, we find that $p_0 = 2(1-q)^{N_I} - (1-q)^{2N_I}$, in agreement with Eq.~(\ref{simple_p0}).  The average number of $nm$th-island heralds per pump pulse can now be evaluated numerically by using Eq.~(\ref{simple_pk}) in Eq.~(\ref{E(H)}).   

\subsection{Chahine~\emph{et al}.\@ Signal-Path Erasure Operation \label{performanceB}}
The work in this section parallels what was done by Shapiro~\emph{et al}.~\cite{Shapiro2025} to derive the results presented in Sec.~\ref{performanceA}.  The jumping-off point in the present section is therefore the anti-normally ordered characteristic function, $\chi^{\rho_{SI}}_A(\bzeta)$, for Fig.~\ref{SagnacSource_fig}'s $SI = S_HS_VI_HI_V$ modes from single islands of its cw and ccw SPDCs~\cite{footnote9}:
\begin{equation}
\chi_A^{\rho_{SI}}(\bzeta) \equiv 
{\rm Tr}\!\left[\hat{\rho}_{SI}e^{-\bzeta^\dagger\bhata}e^{\bhata^\dagger\bzeta}\right] = 
\exp\!\left[-G\bzeta^\dagger\bzeta +2\sqrt{G(G-1)}\,{\rm Re}(\bzeta^T_S\bzeta_I)\right],
\end{equation} 
where $\hat{\rho}_{SI} = |\psi\rangle_{SI}\,{}_{SI}\langle \psi|$, $\bzeta^\dagger \equiv [\begin{array}{cc}\bzeta_S^\dagger & \bzeta_I^\dagger\end{array}]$ and $\bhata^\dagger \equiv [\begin{array}{cc}\bhata_S^\dagger & \bhata_I^\dagger\end{array}]$, with $
\bzeta^\dagger_K \equiv [\begin{array}{cc}\zeta^*_{K_H} & \zeta^*_{K_V}\end{array}]$ and $\bhata_K^\dagger \equiv [\begin{array}{cc}\hat{a}^\dagger_{K_H} & \hat{a}^\dagger_{K_V} \end{array}]$ for $K=S,I$.  To get $\tilde{\rho}_{S\mid \psi^-}$, the unnormalized conditional density operator for the signal modes given a $\psi^-$ herald, we first define the $I' = I'_HI'_V$ modes by
\begin{equation}
\hat{a}_{I'_P} = \sqrt{\eta_T}\,\hat{a}_{I_P} + \sqrt{1-\eta_T}\,\hat{a}_{T_P},\mbox{ for $P = H,V$},
\end{equation} 
where the $\{\hat{a}_{T_P}\}$ modes are in their vacuum states, which gives us
\begin{align}
\chi_A^{\rho_{SI'}}(\bzeta) &= \chi_A^{\rho_{SI}}(\bzeta_S,\sqrt{\eta_T}\,\bzeta_I)\,
{}_T\langle {\bf 0}|e^{-\sqrt{1-\eta_T}\,\bzeta_I^\dagger\bhata_T}e^{\sqrt{1-\eta_T}\,\bhata_T^\dagger \bzeta_I}|{\bf 0}\rangle_T \\[.05in]
&= \exp\!\left[-G\bzeta_S^\dagger\bzeta_S-[\eta_T(G-1) + 1]\bzeta_I^\dagger\bzeta_I +2\sqrt{\eta_TG(G-1)}\,{\rm Re}(\bzeta^T_S\bzeta_I)\right].
\end{align}
The \emph{unnormalized} conditional density operator $\tilde{\rho}_{S\mid \psi^-}$ is then obtained from
\begin{equation}
\tilde{\rho}_{S\mid \psi^-} = {}_{I'_V}\langle 1|\,{}_{I'_H}\langle 1|\hat{\rho}_{SI'}|1\rangle_{I'_H}|1\rangle_{I'_V} = {}_{I'_V}\langle 1|\,{}_{I'_H}\langle 1|\!\left(\int\!\frac{{\rm d}^8\bzeta}{\pi^4}\, 
\chi_A^{\rho_{SI'}}(\bzeta)e^{-\bhata^{\prime\dagger}\bzeta} e^{\bzeta^\dagger\bhata'}\right)|1\rangle_{I'_H}|1\rangle_{I'_V},
\end{equation}
where $|1\rangle_{I'_P}$ for $P=H,V$ is the single-photon Fock state of the $I'_P$ mode, and $\bhata^{\prime\dagger} \equiv [\begin{array}{cccc} \hat{a}^\dagger_{S_H} & \hat{a}^\dagger_{S_V} & \hat{a}^\dagger_{I'_H} & \hat{a}^\dagger_{I'_V} \end{array}]$.  The right-hand side of this equation is evaluated in Appendix~\ref{AppendA} with the result being
\begin{equation}
\tilde{\rho}_{S\mid \psi^-} = \frac{[\eta_T(G-1)]^2}{[\eta_T(G-1)+1]^4}
\int\!\frac{{\rm d}^4\bzeta_S}{\pi^2}\,\chi_A^{\rho_{S\mid\psi^-}}(\bzeta_S)e^{-\bhata_S^\dagger\bzeta_S}e^{\bzeta_S^\dagger\bhata_S},
\end{equation}
where
\begin{equation}
\chi_A^{\rho_{S\mid\psi^-}}(\bzeta_S) = e^{-\bzeta_S^\dagger\bzeta_S/N_S}\left[1-\frac{|\zeta_{S_H}|^2}{N_S}\right]\left[1-\frac{|\zeta_{S_V}|^2}{N_S}\right],
\label{antinorm_Chahine}
\end{equation}
is the anti-normally ordered characteristic function of the \emph{normalized} density operator, $\hat{\rho}_{S\mid \psi^-}$, for the $S$ modes given there has been a $\psi^-$ herald.

At this juncture we immediately find that the probability of a $\psi^-$ herald from the $n$th-island cw SPDC and the $m$th-island ccw SPDC satisfies
\begin{equation}
\Pr(\mathcal{H}_{nm}) = {\rm Tr}(\tilde{\rho}_{S\mid \psi^-}) = \chi_A^{\tilde{\rho}_{S\mid\psi^-}}({\bf 0}) = 
 \frac{[\eta_T(G-1)]^2}{[\eta_T(G-1)+1]^4},
 \label{ChahineHeralding}
\end{equation}
which is the expected result, because the islands' idlers are in iid thermal states with average photon number $\eta_T(G-1)$.  In Sec.~\ref{performanceA}'s notation we have that the Chahine~\emph{et al}.\@ source has $\Pr(\mathcal{H}_{nm}) = \Pr(\mathcal{H}_{H_n})\Pr(\mathcal{H}_{V_m})$ with 
\begin{equation}
\Pr(\mathcal{H}_{H_n}) = \Pr(\mathcal{H}_{V_m}) = \frac{\eta_T(G-1)}{[\eta_T(G-1)+1]^2}.
\end{equation}
Setting $q = \eta_T(G-1)/[\eta_T(G-1)+1]^2$, the derivation in Sec.~\ref{performanceA} now 
gives us the probability, $p_k$, for Chahine~\emph{et al}.'s source to have $k$ SPCI heralds from a single pump pulse.  Knowledge of $p_k$ will enable us to instantiate this source's multiple-memory and single-memory entanglement distribution rates, from Eqs.~(\ref{ReMultiple}) and (\ref{ReSingle}), once we find its Bell-state probability and Bell-state fidelity.  These quantities, and other performance metrics, are derived in Appendix~\ref{AppendA}.  There we find that
\begin{equation}
\Pr({\rm loadable}) = 1-2\tilde{N}_S^2\!\left(1-\frac{\eta_R\tilde{N}_S}{2N_S}\right)^2 + 
N_S^{\prime 2}\!\left(1-\frac{\eta_RN'_S}{N_S}\right)^2, \mbox{ where $\tilde{N}_S \equiv 2N_S'/(N_S'+1)$}, 
\label{PrLoadableChahine}
\end{equation} 
\begin{equation}
\Pr(|\psi^-\rangle_{\tilde{\bf S}}\mid \psi^-) = N_S^{\prime 2}\!\left[\frac{(1-N_S')^2}{2} -\frac{\eta_RN_S'(1-2N_S')(1-N_S')}{N_S} + \frac{\eta_R^2N_S^{\prime 2}(1-2N_S')^2}{2N_S^2}
\right],
\label{PrCorrectChahine}
\end{equation}
\begin{align}
\Pr(|\psi^+\rangle_{\tilde{\bf S}}\mid \psi^-) = 0, 
\end{align}
and
\begin{equation}
\Pr(|\phi^\pm\rangle_{\tilde{\bf S}}\mid \psi^-) = N_S^{\prime 2}\!\left[\frac{(1-N_S')^2}{2} -\frac{\eta_RN_S'(1-2N_S')(1-N_S')}{N_S} + \frac{\eta_R^2N_S^{\prime 2}(1-3N_S')(1-N_S')}{2N_S^2}\right].
\label{PrErrorChahine}
\end{equation}
The preceding results then give us
\begin{align}
&\Pr({\rm Bell}) = \nonumber \\[.05in]
&\,N_S^{\prime 2}\!\left[\frac{3(1-N_S')^2}{2} - \frac{3\eta_RN_S'(1-2N_S')(1-N_S')}{N_S} + \frac{\eta_R^2N_S^{\prime 2}[(1-2N_S')^2+ 2(1-3N_S')(1-N_S')]}{2N_S^2}\right],
\label{BellChahine}
\end{align}
$\mathcal{B} = \Pr({\rm Bell})/\Pr({\rm loadable})$, and $\mathcal{F} = \Pr(|\psi^-\rangle_{\tilde{\bf S}}\mid \psi^-)/\Pr({\rm Bell})$.  

Appendix~A also shows that, as was the case for islands-based ZALM, $\hat{\rho}^{(\mathcal{H_B})}_{\tilde{S}\mid \psi^-}$ is diagonal in the Bell basis, hence easily expressible from Eqs.~(\ref{PrCorrectChahine})--(\ref{PrErrorChahine}).  However, unlike the case for islands-based ZALM, $\hat{\rho}^{(\mathcal{H_B})}_{\tilde{S}\mid \psi^-}$ is \emph{not} a Werner state, because $\Pr(|\psi^+\rangle_{\tilde{S}}\mid \psi^-) = 0$.  It is also worth noting that, as was the case for islands-based ZALM, Chahine \emph{et al}.'s  signal-path erasure scheme gives ideal, propagation-loss-limited performance when $\eta_T = 1$, i.e.,
\begin{equation}
\Pr({\rm loadable})  = \Pr({\rm Bell}) = \Pr(|\psi^-\rangle_{\tilde{S}}\mid \psi^-) = \eta_R^2/2.
\end{equation}

\subsection{Unheralded Operation \label{performanceC}}
Our last configuration for which we seek performance metrics is unheralded operation.  Once again, we will begin our metrics quest from the relevant anti-normally ordered characteristic function.  Here, because unheralded operation's Sagnac source in Fig.~\ref{unheralded_fig} differs from Chahine~\emph{et al}.'s source in Fig.~\ref{SagnacSource_fig} only by the addition of a HWP to the idler's output path, we can obtain the former's $\chi_A^{\rho_{SI}}(\bzeta)$ by applying the HWP to the idler portion of the latter's $\chi_A^{\rho_{SI}}(\bzeta)$.   What results is 
\begin{equation}
\left.\chi_A^{\rho_{SI}}(\bzeta)\right|_{\rm unheralded} = \left.\chi_A^{\rho_{SI}}(\bzeta')\right|_{\rm Chahine},
\end{equation}
where $\bzeta'^\dagger  \equiv [\begin{array}{cc}\bzeta_S^\dagger & \bzeta_I^{\prime\dagger} \end{array}]$ with $\bzeta_I^{\prime\dagger} \equiv [\begin{array}{cc}\zeta_{I_V}^* & -\zeta_{I_H}^*\end{array}]$, which gives the unheralded source's characteristic function as
\begin{equation}
\chi_A^{\rho_{SI}}(\bzeta) = \exp\!\left[-G\bzeta^\dagger\bzeta +2\sqrt{G(G-1)}\,{\rm Re}(\bzeta^T_S\Omega\bzeta_I)\right],
\label{unheralded_chiA}
\end{equation}
with 
\begin{equation}
\Omega \equiv \left[\begin{array}{cc} 0 & 1\\[.05in] -1 & 0 \end{array}\right].
\end{equation}
Accounting for propagation loss, as we have done for the heralded sources, we get
\begin{equation}
\chi_A^{\rho_{S_AI_B}}(\bzeta) = \exp\!\left[-[\eta_R(G-1)+1]\bzeta^\dagger\bzeta +2\eta_R\sqrt{G(G-1)}\,{\rm Re}(\bzeta^T_S\Omega\bzeta_I)\right],
\label{unheralded_chiAloss}
\end{equation}
for the anti-normally ordered characteristic function for the light arriving at Alice and Bob's QRXs.

With the preceding characteristic function in hand, we could now proceed to find the metrics of interest.  Instead, we postpone our metrics finding to fulfill a promise made in Sec.~\ref{configurationC}, namely, demonstrating that the unheralded source's state resulting from a single pump pulse is given by Eq.~(\ref{unheralded_islands_state}).  For this purpose, it suffices to match $\hat{\rho}_{SI} \equiv |\psi\rangle_{SI}\,{}_{SI}\langle \psi|$ with
\begin{equation}
|\psi\rangle_{SI} = \sum_{m=0}^\infty\sqrt{P_m}\,|m\rangle_{S_H}|m\rangle_{I_V}\sum_{m'=0}^\infty(-1)^{m^\prime}\sqrt{P_{m^\prime}}\,|m'\rangle_{S_V}|m'\rangle_{I_H},
\label{numberstate_unheralded}
\end{equation}
to $\hat{\rho}_{SI}$ from 
\begin{equation}
\hat{\rho}_{SI} = \int\!\frac{{\rm d}^8\bzeta}{\pi^4}\,\exp\!\left[-G\bzeta^\dagger\bzeta +2\sqrt{G(G-1)}\,{\rm Re}(\bzeta^T_S\Omega\bzeta_I)\right]e^{-\bhata^\dagger\bzeta}e^{\bzeta^\dagger\bhata},
\label{4D_inverse_transform}
\end{equation}
where $\bhata^\dagger = [\begin{array}{cccc}\hat{a}^\dagger_{S_H}& \hat{a}^\dagger_{S_V} &\hat{a}^\dagger_{I_H} & \hat{a}^\dagger_{I_V}\end{array}]$.

It is well known that the density operator $\hat{\rho}_{SI}$ is completely characterized by its normally-ordered form,
\begin{equation}
\rho^{(n)}_{SI}(\balpha_S,\balpha_I) \equiv {}_I\langle \balpha_I|\,{}_S\langle \balpha_S|\hat{\rho}_{SI}|\balpha_S\rangle_S|\balpha_I\rangle_I,
\end{equation}
where $|\balpha_K\rangle_K \equiv |\alpha_{K_H}\rangle_{K_H}|\alpha_{K_V}\rangle_{K_V}$ for $K = S,I$ are two-mode coherent states.  From Eq.~(\ref{numberstate_unheralded}) we have that
\begin{align}
\rho^{(n)}_{SI}(\balpha_S,\balpha_I) &= \left|{}_I\langle \balpha_I|\,{}_S\langle \balpha_S|\sum_{m=0}^\infty\sqrt{P_m}\,|m\rangle_{S_H}|m\rangle_{I_V}\sum_{m'=0}^\infty(-1)^{m^\prime}\sqrt{P_{m^\prime}}\,|m'\rangle_{S_V}|m'\rangle_{I_H}|\balpha_S\rangle_S|\balpha_I\rangle_I\right|^2 \\[.05in]&=\frac{\exp\!\left[-\balpha^\dagger\balpha + 2\sqrt{\frac{G-1}{G}}\,{\rm Re}(\balpha_S^T\Omega\balpha_I)\right]}{G^2},
\end{align}
where $\balpha^\dagger \equiv [\begin{array}{cc} \balpha_S^\dagger & \balpha_I^\dagger \end{array}]$ with $\balpha_K^\dagger \equiv [\begin{array}{cc} \alpha_{K_H}^* & \alpha_{K_V}^* \end{array}]$ for $K = S,I$.
On the other hand, the transform-domain approach gives
\begin{align}
 \rho^{(n)}_{SI}&(\balpha_S,\balpha_I)  \nonumber \\[.05in]
&= \int\!\frac{{\rm d}^8\bzeta}{\pi^4}\,\exp\!\left[-G\bzeta^\dagger\bzeta +2\sqrt{G(G-1)}\,{\rm Re}(\bzeta^T_S\Omega\bzeta_I)\right]{}_I\langle \balpha_I|\,{}_S\langle \balpha_S|e^{-\bhata^\dagger\bzeta}e^{\bzeta^\dagger\bhata}
 |\balpha_S\rangle_S|\balpha_I\rangle_I \\[.05in]
 &= \int\!\frac{{\rm d}^8\bzeta}{\pi^4}\,\exp\!\left[-G\bzeta^\dagger\bzeta +2\sqrt{G(G-1)}\,{\rm Re}(\bzeta^T_S\Omega\bzeta_I)\right]e^{-\balpha^\dagger\bzeta}e^{\bzeta^\dagger\balpha} \\[.05in]
 &= \frac{\exp\!\left[-\balpha^\dagger\balpha + 2\sqrt{\frac{G-1}{G}}\,{\rm Re}(\balpha_S^T\Omega\balpha_I)\right]}{G^2},
 \end{align}
 and our demonstration is complete.

Appendix~\ref{AppendB} provides the details for obtaining unheralded operation's performance, viz.,
\begin{equation}
\Pr({\rm loadable}) = 1 - \frac{2}{[\eta_R(G-1) + 1]^2} + \frac{1}{[\eta_R(2-\eta_R)(G-1) + 1]^2},
\label{PrLoadableUnheralded}
\end{equation}
\begin{equation}
\Pr(|\psi^-\rangle_{S_AI_B}) = \frac{\eta_R^2(G-1)[3G-1+\eta_R(\eta_R-2)(G-1)]} {[\eta_R(\eta_R-2)(G-1) -1]^4},
\label{PrCorrectUnheralded}
\end{equation}
\begin{equation}
\Pr(|\psi^+\rangle_{S_AI_B})=  \Pr(|\phi^\pm\rangle_{S_AI_B}) = \frac{[\eta_R(\eta_R-1)(G-1)]^2} {[\eta_R(\eta_R-2)(G-1) -1]^4} 
\label{PrErrorUnheralded}
\end{equation}
\begin{equation}
\Pr({\rm Bell}) = \frac{\eta_R^2(G-1)[3G-1+\eta_R(\eta_R-2)(G-1)] + 3[\eta_R(\eta_R-1)(G-1)]^2}{[\eta_R(\eta_R-2)(G-1) -1]^4},
\label{BellUnheralded}
\end{equation}
$\mathcal{B} = \Pr({\rm Bell})/\Pr({\rm loadable})$, and $\mathcal{F} = \Pr(|\psi^ -\rangle_{S_AI_B})/\Pr({\rm Bell})$.  Appendix~\ref{AppendB} also shows that $\hat{\rho}^{(\mathcal{H}_B)}_{S_AI_B}$ is diagonal in the Bell basis, hence easily written down from Eqs.~(\ref{PrCorrectUnheralded}) and (\ref{PrErrorUnheralded}).  Moreover,  as was the case for islands-based ZALM, $\hat{\rho}^{(\mathcal{H}_B)}_{S_AI_B}$ is a Werner state.

\subsection{Performance Comparisons \label{performanceD}}
\begin{figure}[hbt]
    \centering
\includegraphics[width=3.5in]{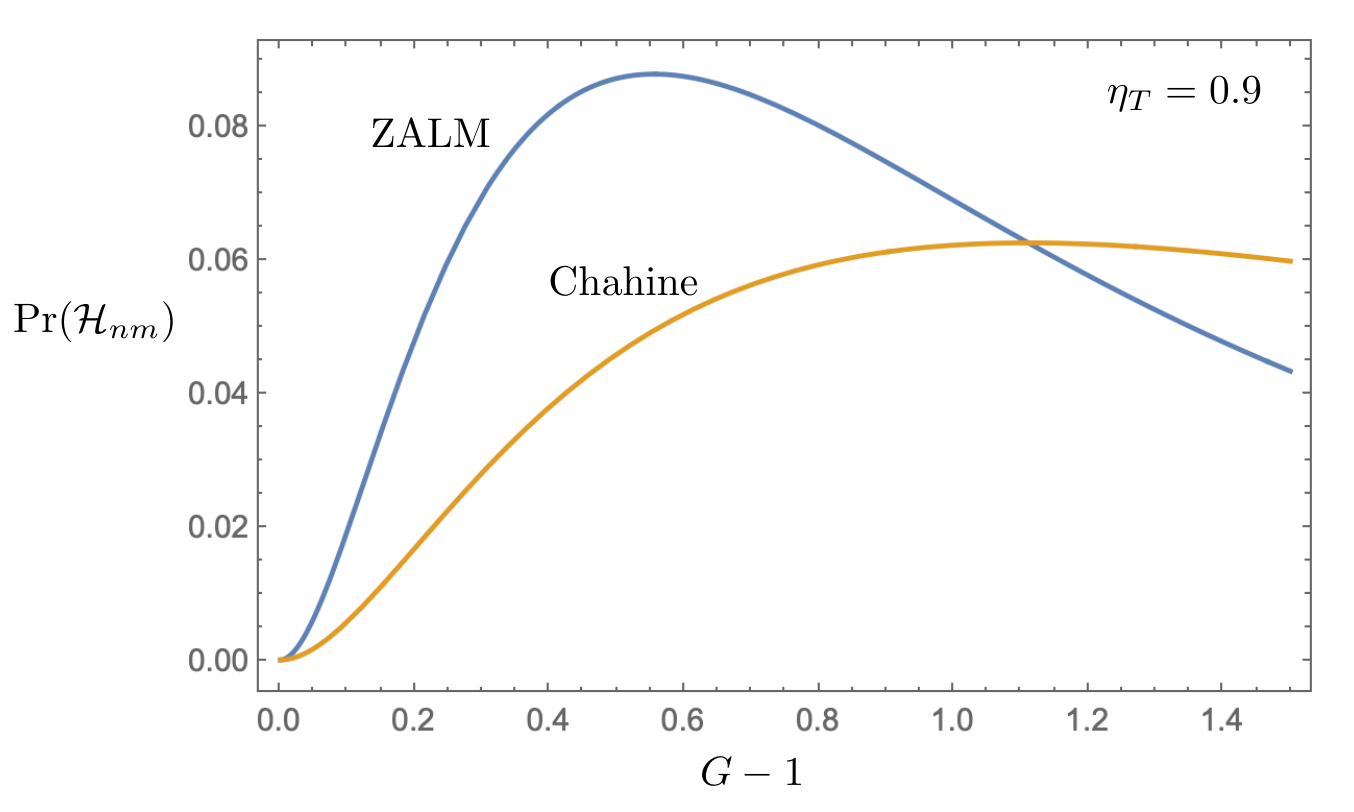}
\caption{Plots, versus the average number of signal-idler photon pairs per SPDC per pump pulse (G-1), of the $nm$th-island heralding probabilities $\Pr(\mathcal{H}_{nm})$ from Eq.~(\ref{ZALMheralding}) (for ZALM), and from Eq.~(\ref{ChahineHeralding}) (for Chahine \emph{et al}.'s  source) assuming  90\% efficient QTXs ($\eta_T = 0.9$).  \label{HeraldingProbs_fig}}    
\end{figure}

We are now ready to address the issue raised in this paper's title:  ``To herald or not to herald''.  We begin with some comparisons between islands-based ZALM and Chahine~\emph{et al}.'s signal-path erasure source.  Figure~\ref{HeraldingProbs_fig} plots, versus the average number of signal-idler photon pairs per SPDC per pump pulse $(G-1)$, their $nm$th-island heralding probabilities, $\Pr(\mathcal{H}_{nm})$, from Eq.~(\ref{ZALMheralding}) (for ZALM) and Eq.~(\ref{ChahineHeralding}) (for Chahine \emph{et al}.'s  source) assuming 90\% efficient QTXs ($\eta_T = 0.9$).  We see that ZALM's dual-Sagnac source outperforms Chahine~\emph{et al}.'s single-Sagnac configuration for this figure of merit at subunit $G-1$, but keep in mind that this paper's goal is maximizing the per-pump-pulse rate of entanglement distribution.  So let's continue working through our other metrics en route to that ultimate comparison.  

\begin{figure}[hbt]
    \centering
\includegraphics[width=3.5in]{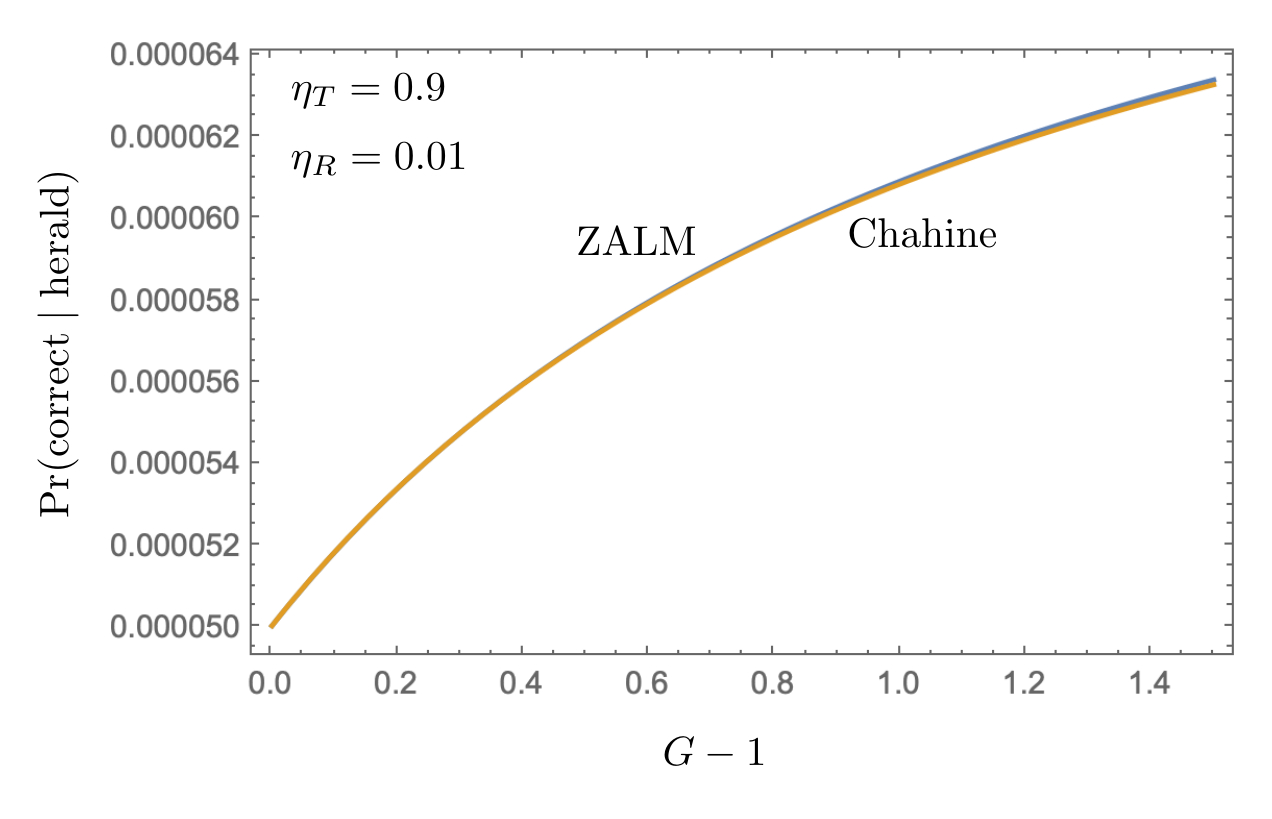}
\caption{Plots, versus the average number of signal-idler photon pairs per SPDC per pump pulse (G-1), of $\Pr(\mbox{correct} \mid \mbox{herald})$, the probability that the heralded state reaches Alice and Bob's QRXs, for ZALM and Chahine \emph{et al}.'s  source~\cite{footnote10}.  The QTXs are assumed 90\% efficient ($\eta_T = 0.9$) and each QTX-to-QRX connection has 20\,dB propagation loss ($\eta_R = 0.01$).  \label{PrCorrectHeralded_fig}}    
\end{figure}

\begin{figure}[hbt]
    \centering
\includegraphics[width=3.5in]{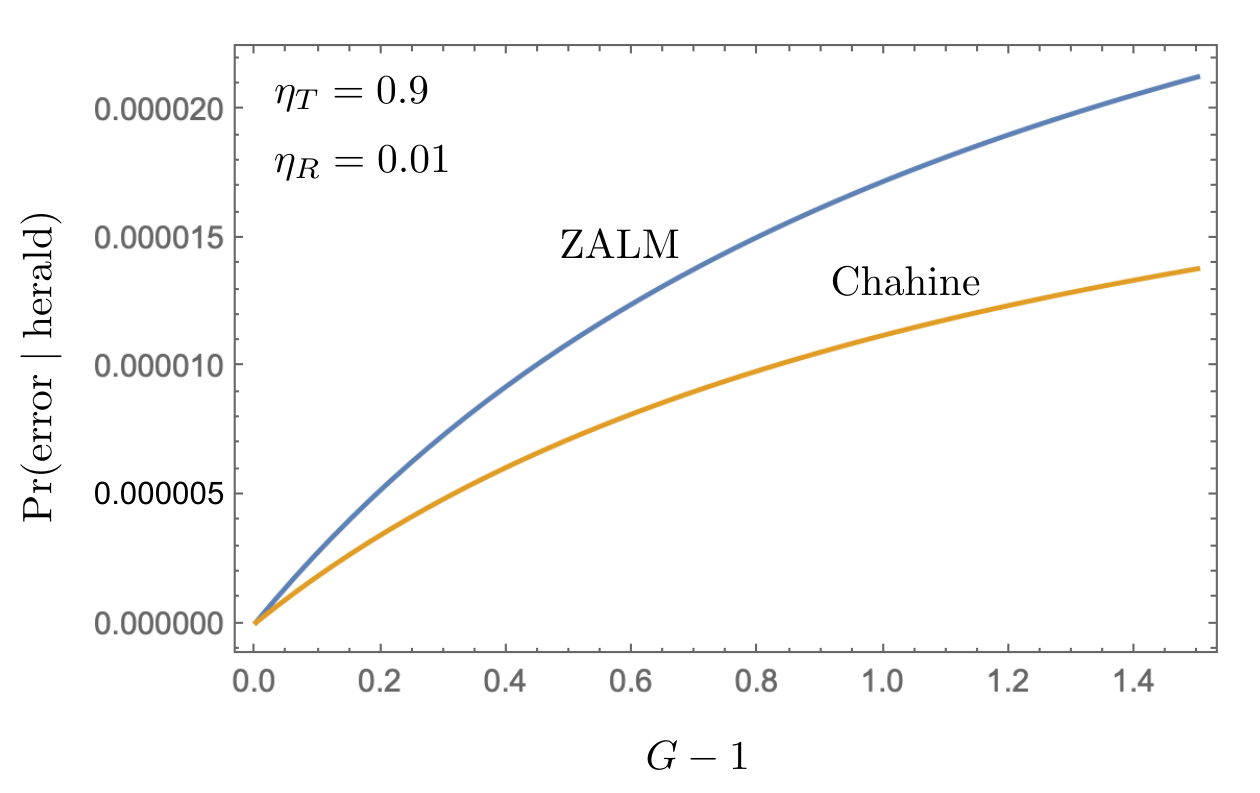}
\caption{Plots, versus the average number of signal-idler photon pairs per SPDC per pump pulse $(G-1)$, of $\Pr(\mbox{error} \mid \mbox{herald})$, the probability that an erroneous Bell state reaches Alice and Bob's QRXs, given there has been a herald, for ZALM and Chahine \emph{et al}.'s  source~\cite{footnote11}.  The QTXs are assumed 90\% efficient ($\eta_T = 0.9$) and each QTX-to-QRX connection has 20\,dB propagation loss ($\eta_R = 0.01$).  \label{PrErrorHeralded_fig}}    
\end{figure}

Figure~\ref{PrCorrectHeralded_fig} compares the two heralded sources' probabilities for the heralded state to reach Alice and Bob's QRXs when $\eta_T = 0.9$ and $\eta_R = 0.01$.  Here we see virtually no difference in their performance.  The situation is different, however, when we consider $\Pr(\mbox{error}\mid\mbox{herald})$, which is the probability that an \emph{incorrect} Bell state is received given there has been a herald.  As shown in Fig.~\ref{PrErrorHeralded_fig}, the ZALM source's error probability increases more rapidly than does that of Chahine~\emph{et al}.'s source, as the average number of signal-idler photon pairs per SPDC per pump pulse increases.  This trend indicates that the ZALM source is more susceptible to the ill-effects of multipair events than is Chahine~\emph{et al}.'s   

\begin{figure}[hbt]
    \centering
\includegraphics[width=3.5in]{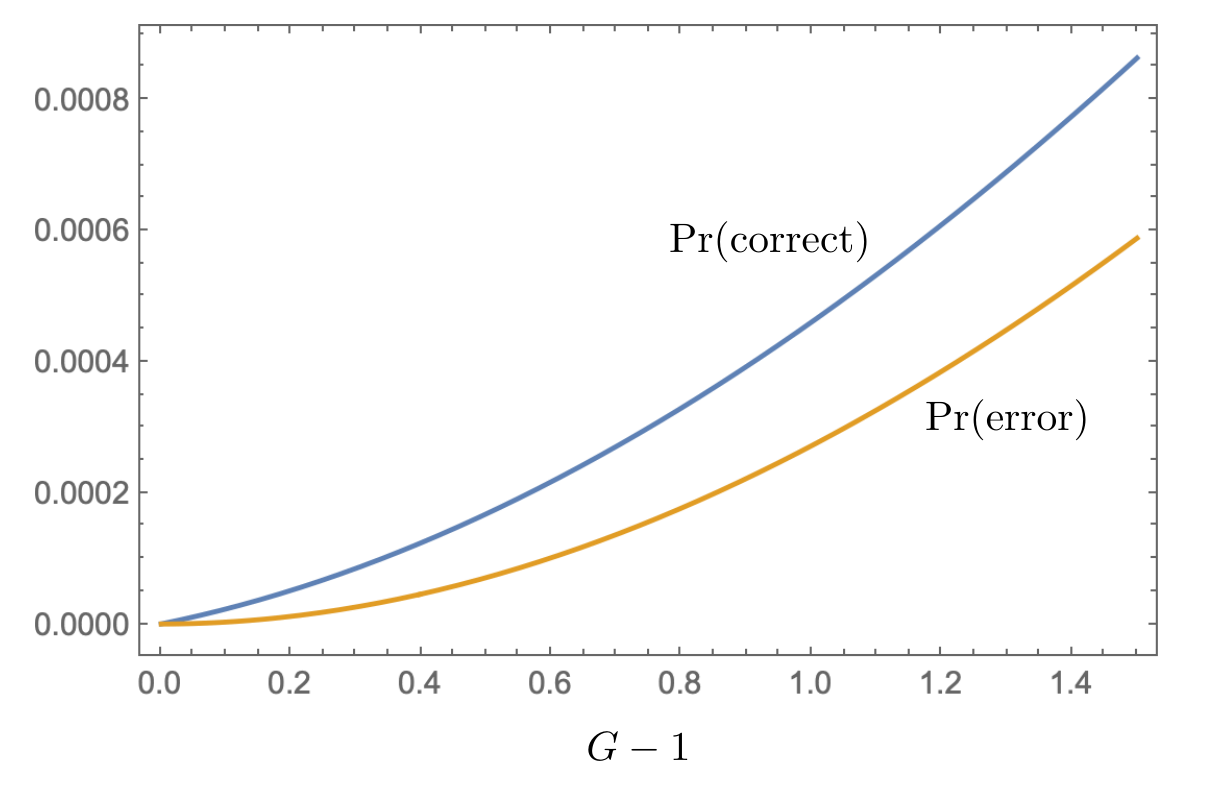}
\caption{Plots, versus the average number of signal-idler photon pairs per SPDC per pump pulse $(G-1)$, of $\Pr(\mbox{correct})$, the probability that the correct ($\psi^-$) Bell state reaches Alice and Bob's QRXs, and $\Pr(\mbox{error})$, the probability that an erroneous ($\psi^+,\phi^+$, or $\phi^-$) Bell state arrives for the unheralded source~\cite{footnote12}.  Each QTX-to-QRX connection has 20\,dB propagation loss ($\eta_R = 0.01$)  \label{PrCorrectErrorUnheralded_fig}}    
\end{figure}

It is now time to start comparing our heralded sources to unheralded operation.  Toward that end Fig.~\ref{PrCorrectErrorUnheralded_fig} plots the unheralded source's $\Pr(\mbox{correct})$ and $\Pr(\mbox{error})$ versus $G-1$ for $\eta_R = 0.01$.  In comparison with the heralded sources' plots, the unheralded case's probabilities of correct and erroneous reception are much higher and more rapidly increasing with increasing $G-1$ over the range being plotted.  These differences are due to the heralded sources' PNR-capable SPDs strongly suppressing multipair events when operated at high (here 90\%) quantum efficiency.  The presence of multipair events in the light transmitted by the unheralded source results in increasing numbers of Bell states arriving at the QRXs after propagation through 20-dB-loss media as $G-1$ increases.  It remains to be seen whether these increases are or are not beneficial to the entanglement distribution rate.  The next step in that direction is to compare our three configurations' Bell-state fractions and Bell-state fidelities. 

\begin{figure}[hbt]
    \centering
\includegraphics[width=3.5in]{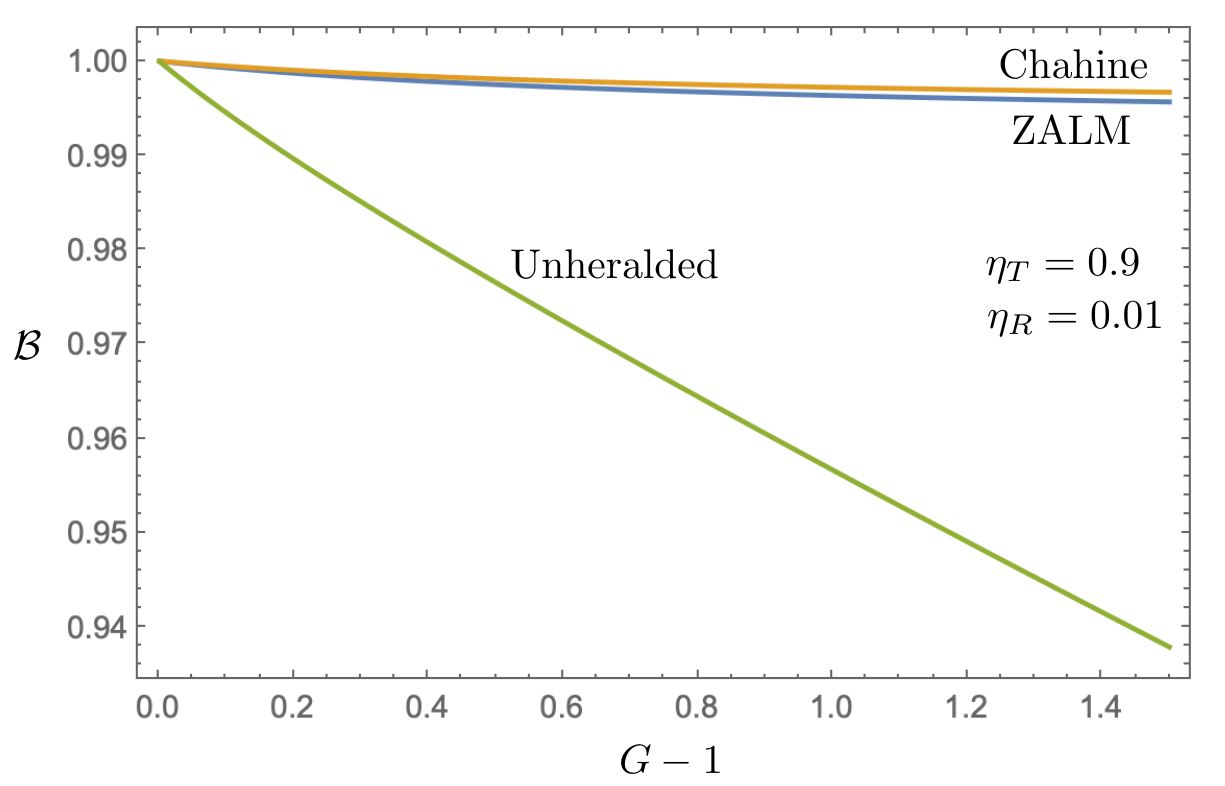}
\caption{Plots, versus the average number of signal-idler photon pairs per SPDC per pump pulse $(G-1)$, of the Bell-state fractions for the ZALM source, Chahine~\emph{et al}.'s signal-path erasure source, and unheralded operation.  The heralded sources' have 90\% QTX efficiencies, $\eta_T = 0.9$.  All three sources' QTX-to-QRX connections have 20\,dB propagation loss ($\eta_R = 0.01$)  \label{BellFraction-comp_fig}}    
\end{figure}

\begin{figure}[hbt]
    \centering
\includegraphics[width=3.5in]{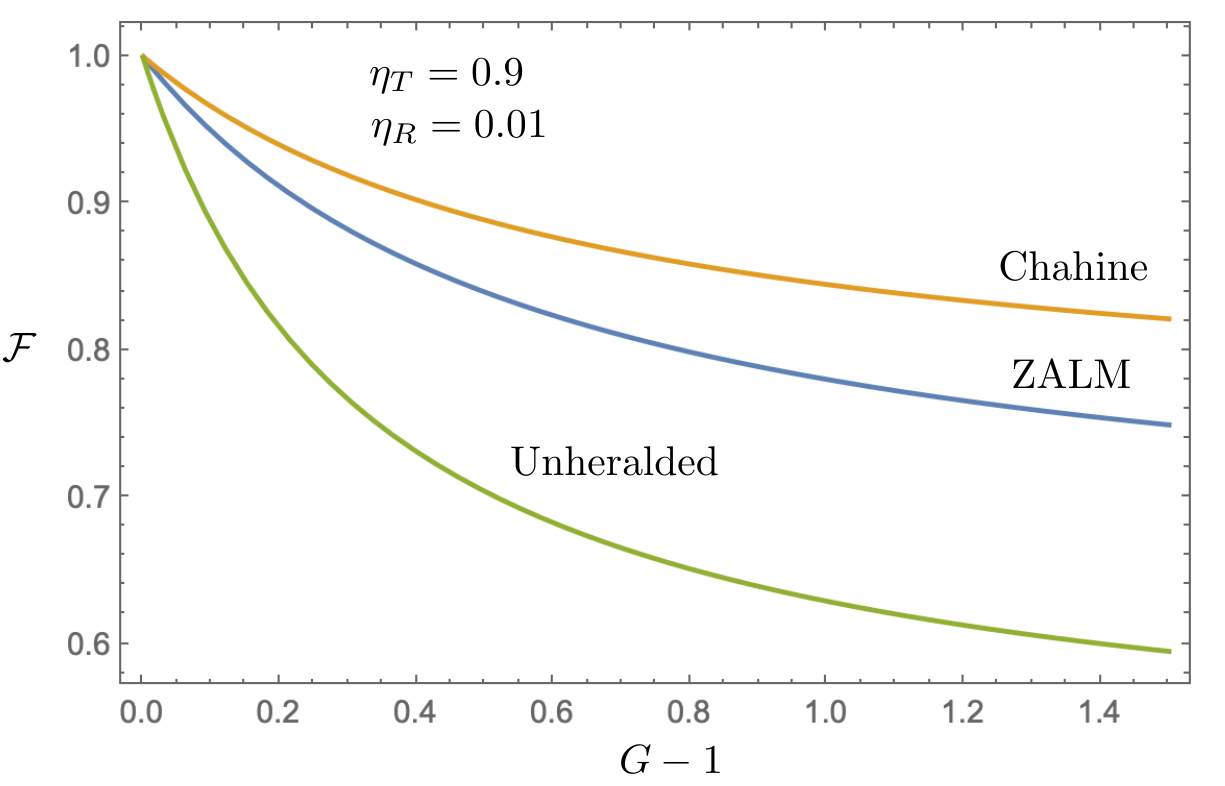}
\caption{Plots, versus the average number of signal-idler photon pairs per SPDC per pump pulse $(G-1)$, of the Bell-state fidelities for the ZALM source, Chahine~\emph{et al}.'s signal-path erasure source, and unheralded operation.  The heralded sources' have 90\% QTX efficiencies, $\eta_T = 0.9$.  All three sources' QTX-to-QRX connections have 20\,dB propagation loss ($\eta_R = 0.01$)  \label{BellFidelity-comp_fig}}    
\end{figure}

Figures~\ref{BellFraction-comp_fig} and \ref{BellFidelity-comp_fig} plot our three configurations' $\mathcal{B}$ and $\mathcal{F}$ behaviors versus $G-1$ where the heralded systems' QTX efficiencies are both 90\% and all three systems suffer 20\,dB QTX-to-QRX propagation loss.  These figures show the deleterious effect of the increased multipair emissions that  accompanies increasing $G-1$.  The heralded sources' Bell-state fractions are relatively immune to multipairs in comparison with that of the unheralded source, but they all suffer very significant fidelity losses.  Note that Chahine~\emph{et al}.'s configuration outperforms ZALM in both its Bell-state fraction and Bell-state fidelity.  
 

Figures.~\ref{HeraldingProbs_fig}--\ref{BellFidelity-comp_fig} present a mixed message insofar as which configuration is best.  In particular, ZALM has the advantage over the Chahine~\emph{et al}.~source in heralding probability, but ZALM is at a slight disadvantage with respect to the Chahine~\emph{et al}.~source in Bell-state fraction and at a more significant disadvantage in Bell-state fidelity.  Moreover, both heralded systems outperform unheralded operation in both those metrics, owing to their heraldings' suppression of multipair events.  The decisive comparison, if we ignore the complexity disparities of their source and mode-conversion implementations, is their entanglement distribution rates.  

A comprehensive exploration of the parameter space determining those rates is beyond the scope of this paper.  Instead, in the rest of this section, we will compare the distribution rates for both single herald and multiple herald operation.  Shapiro~\emph{et al}.~\cite{Shapiro2025} showed for $\eta_T = 0.9$ and $\eta_R = 0.01$ that ZALM with $G-1 = 0.0173$ achieved $\mathcal{B}\ge 0.9998$ and $\mathcal{F}=0.99$.  Furthermore, with $\mathcal{R}_P = 10^9\,{\rm s}^{-1}$, $N_I = 41$, and single-herald operation, that ZALM system's entanglement distribution rate was $\mathcal{R}_E = 2.54 \times 10^4\,{\rm s}^{-1}$, while its entanglement generation probability, $\Pr({\rm ent}) = 0.250$, reached the quasideterministic threshold.  We will use those results as a starting point for our distribution-rate comparison.

\begin{table}
\centering
\begin{tabular}{|c||c|c|c|}\hline
Configuration & $\Delta G_{\rm max}$ for $\mathcal{B} \ge 0.9998$ & $\Delta G_{\rm max}$ for $\mathcal{F} \ge  0.99$ & $\Delta G_{\rm max}$ for both\\[.05in] \hline
ZALM & 0.0258 & 0.0173 & 0.0173 \\[.05in]
Chahine & 0.0347 & 0.0263 & 0.0263 \\[.05in]
Unheralded & 0.00337 & 0.00694 & 0.00337\\[.05in]\hline
\end{tabular}
\caption{Maximum average number of signal-idler pairs per SPDC per pump pulse, $\Delta G_{\rm max} \equiv {\rm max}(G-1)$, that gives $\mathcal{B} \ge 0.9998$, that gives $\mathcal{F} \ge 0.99$, and that satisfies both constraints for ZALM, the Chahine~\emph{et al}.\@ source, and unheralded operation.  These values assume $\eta_T= 0.9$ and $\eta_R = 0.01$.  
\label{dGvalues}}
\end{table}
Table~\ref{dGvalues} lists our three configurations' maximum $G-1$ values, $\Delta G_{\rm max}$, that give $\mathcal{B} \ge 0.9998$, that give $\mathcal{F} \ge 0.99$, and that satisfy both constraints, assuming $\eta_T = 0.9$ and $\eta_R = 0.01$. It is interesting to note that the Bell-fidelity constraint is more severe than the Bell-fraction constraint for the heralded sources, but the reverse is true for unheralded operation.  Also note that unheralded operation requires $G-1 < 0.01$, its oft-employed rule for multipair effects to be negligible, to meet the $\mathcal{B}$ and $\mathcal{F}$ constraints, whereas both heralded schemes can satisfy them while violating that rule.  Figures~\ref{BellFraction-comp_fig} and \ref{BellFidelity-comp_fig} show that $\mathcal{B}$ is a weaker function of $G-1$ than is $\mathcal{F}$ in the $G-1 \ll 1$ regime, \emph{and} increasing $G-1$ in that regime will increase the entanglement distribution rate. Thus we shall back off our goal for the Bell-state fraction to $\mathcal{B} \ge 0.999$ while keeping $\eta_T = 0.9$ and $\eta_R = 0.01$.  Table~\ref{dGvaluesNew} then shows that  all three configurations have their $\Delta G_{\rm max}$ values limited by the Bell-fidelity constraint.  Consequently, unheralded operation's $\Delta G_{\rm max}$ value that satisfies both constraints has nearly doubled, whereas those for the heralded sources are unchanged.  

\begin{table}
\centering
\begin{tabular}{|c||c|c|c|}\hline
Configuration & $\Delta G_{\rm max}$ for $\mathcal{B} \ge 0.999$ & $\Delta G_{\rm max}$ for $\mathcal{F} \ge  0.99$ & $\Delta G_{\rm max}$ for both\\[.05in] \hline
ZALM & 0.148 & 0.0173 & 0.0173 \\[.05in]
Chahine & 0.208 & 0.0263 & 0.0263 \\[.05in]
Unheralded & 0.0171 & 0.00694 & 0.00694\\[.05in]\hline
\end{tabular}
\caption{Maximum average number of signal-idler pairs per SPDC per pump pulse, $\Delta G_{\rm max} \equiv {\rm max}(G-1)$, that gives $\mathcal{B} \ge 0.999$, that gives $\mathcal{F} \ge 0.99$, and that satisfies both constraints for ZALM, the Chahine~\emph{et al}.\@ source, and unheralded operation.  These values assume $\eta_T= 0.9$ and $\eta_R = 0.01$.
\label{dGvaluesNew}}
\end{table}

\begin{figure}[hbt]
    \centering
\includegraphics[width=3.5in]{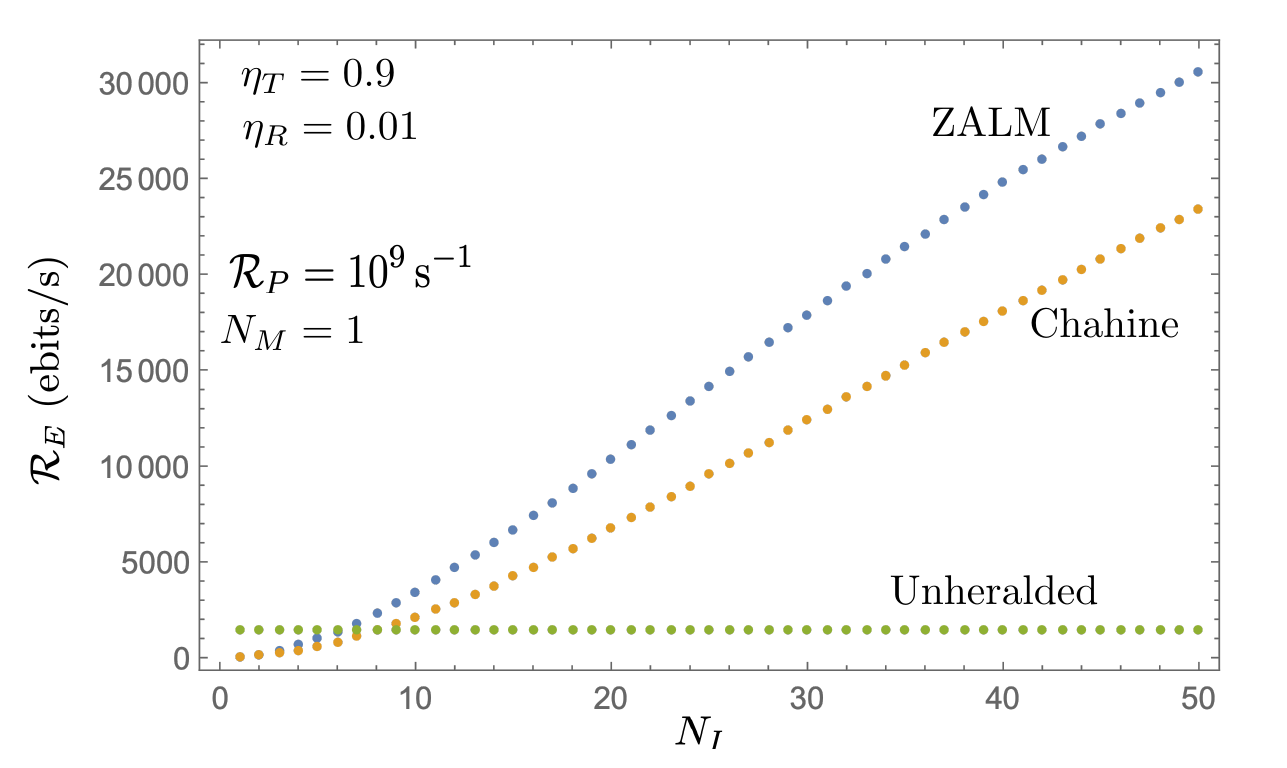}
\caption{Plots of the entanglement distribution rates, $\mathcal{R}_E$, in ebits/s, when Alice and Bob have only a single memory ($N_M=1$) allocated to each pump pulse, assuming $\mathcal{R}_P = 10^9\,{\rm s}^{-1}$, $\eta_T = 0.9$, and $\eta_R = 0.01$.  The $G-1$ values for all three configurations are given by the rightmost column in Table~\ref{dGvaluesNew}.  Unheralded operation's source has a single phase-matched spectral island,  but the heralded sources have $N_I$ islands. 
\label{ReSingleMemory_fig}}    
\end{figure}

Figure~\ref{ReSingleMemory_fig} compares the entanglement distribution rates, $\mathcal{R}_E$, in ebits/s, for our three configurations when Alice and Bob have only a single memory ($N_M=1$) allocated to each pump pulse. The plots assume $\mathcal{R}_P = 10^9\,{\rm s}^{-1}$, $\eta_T = 0.9$, and $\eta_R = 0.01$, and their $G-1$ values ensure that $\mathcal{B} \ge 0.999$ and $\mathcal{F} = 0.99$.  Here, the unheralded source has only a single phase-matched spectral island, while the heralded sources have $N_I$ islands.  As expected, heralded operation greatly outperforms unheralded operation for $N_I \ge 10$, because of the former's multiplexing advantage.  We also see that ZALM outperforms the Chahine~\emph{et al}.\@ source, because its heralding probability advantage overcomes its $\mathcal{B}$ and $\mathcal{F}$ disadvantages.  Note that, for the $G-1$ values in Table~\ref{dGvaluesNew} and $N_I = 50$, our three configurations' $N_M =1$ entanglement generation probabilities are
\begin{equation}
\Pr({\rm ent}) = \left\{\begin{array}{ll}
0.301, & \mbox{for ZALM},\\[.05in]
0.229, & \mbox{for Chahine~\emph{et al}.} \\[.05in]
0.0136, & \mbox{for unheralded operation,} \end{array}\right.
\end{equation}
so only ZALM exceeds the quasideterministic threshold.

\begin{figure}[hbt]
    \centering
\includegraphics[width=3.5in]{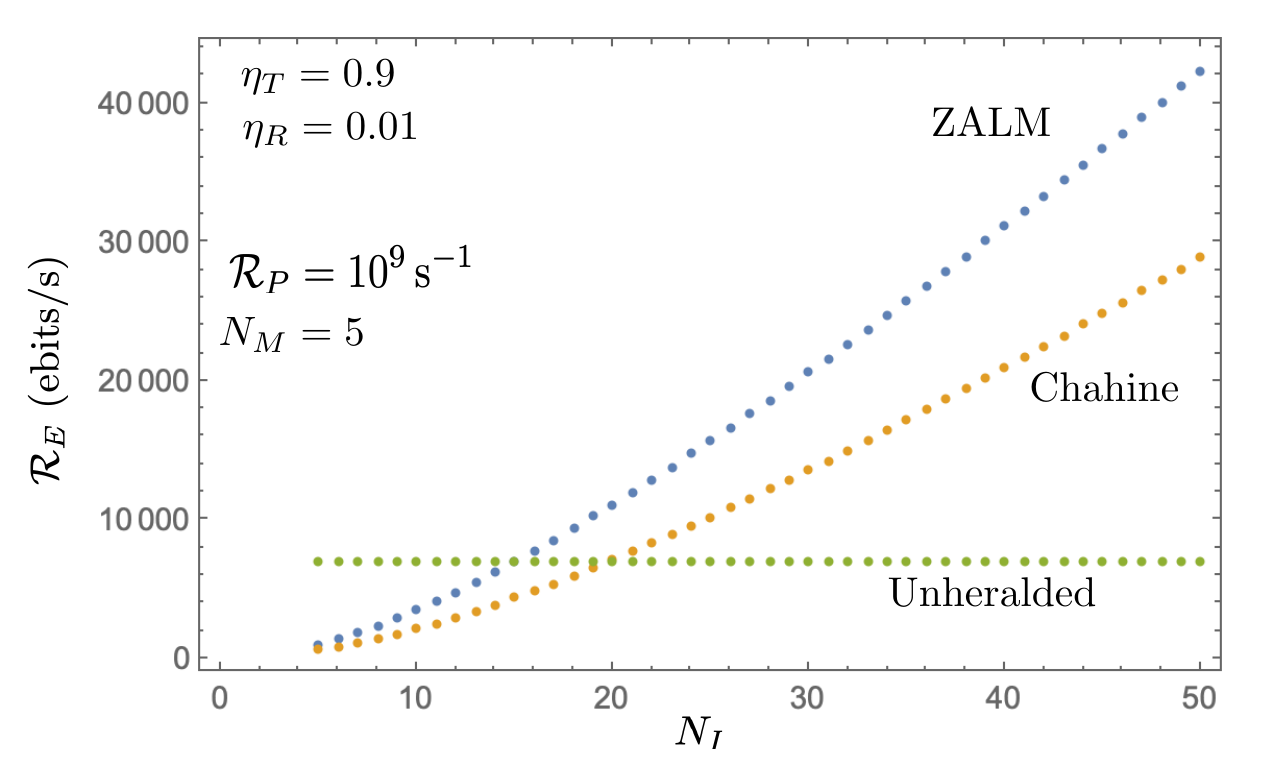}
\caption{Plots of the entanglement distribution rates, $\mathcal{R}_E$, in ebits/s, when Alice and Bob have 5 memories ($N_M=5$) allocated to each pump pulse, assuming $\mathcal{R}_P = 10^9\,{\rm s}^{-1}$, $\eta_T = 0.9$, and $\eta_R = 0.01$.  The $G-1$ values for all three configurations are given by the rightmost column in Table~\ref{dGvaluesNew}.  Unheralded operation's source had $N_M$ phase-matched spectral islands,  while the heralded sources have $N_I$ islands. 
\label{Re5Memories_fig}}    
\end{figure}

Initial (call them first-generation) demonstrations of ZALM or Chahine~\emph{et al}.'s signal-path erasure source are unlikely to have many more than Morrison~\emph{et al}.'s 8 phase-matched spectral islands, and those demonstrations' quantum memory resources, whether for Duan-Kimble or emissive loading, will probably be limited to $N_M = 1$.  Figure~\ref{ReSingleMemory_fig},  which assumes $\mathcal{R}_P = 10^9\,{\rm s}^{-1}$, $\eta_T = 0.9$, $\eta_R = 0.01$, and $G-1$ ensuring that $\mathcal{B} \ge 0.999$ and $\mathcal{F} = 0.99$, then indicates that these will be proof-of-principle demonstrations, rather than greatly advantageous implementations.  Second-generation versions of these multiplexing schemes might offer 10s of islands and a few memories per pump pulse.  Figure~\ref{Re5Memories_fig} illustrates this regime's entanglement distribution rates for Fig.~\ref{ReSingleMemory_fig}'s source and propagation parameters but with $N_M=5$.  Here the unheralded source has $N_M$ phase-matched spectral islands, while the heralded sources continue to have $N_I$ islands. Figure~\ref{Re5Memories_fig}'s lesson is the same as Fig.~\ref{ReSingleMemory_fig}'s in that ZALM outperforms the Chahine~\emph{et al}.\@ source, but here, with enough phase-matched spectral islands, ZALM offers a significant distribution rate advantage over unheralded operation.  

\begin{figure}[hbt]
    \centering
\includegraphics[width=3.5in]{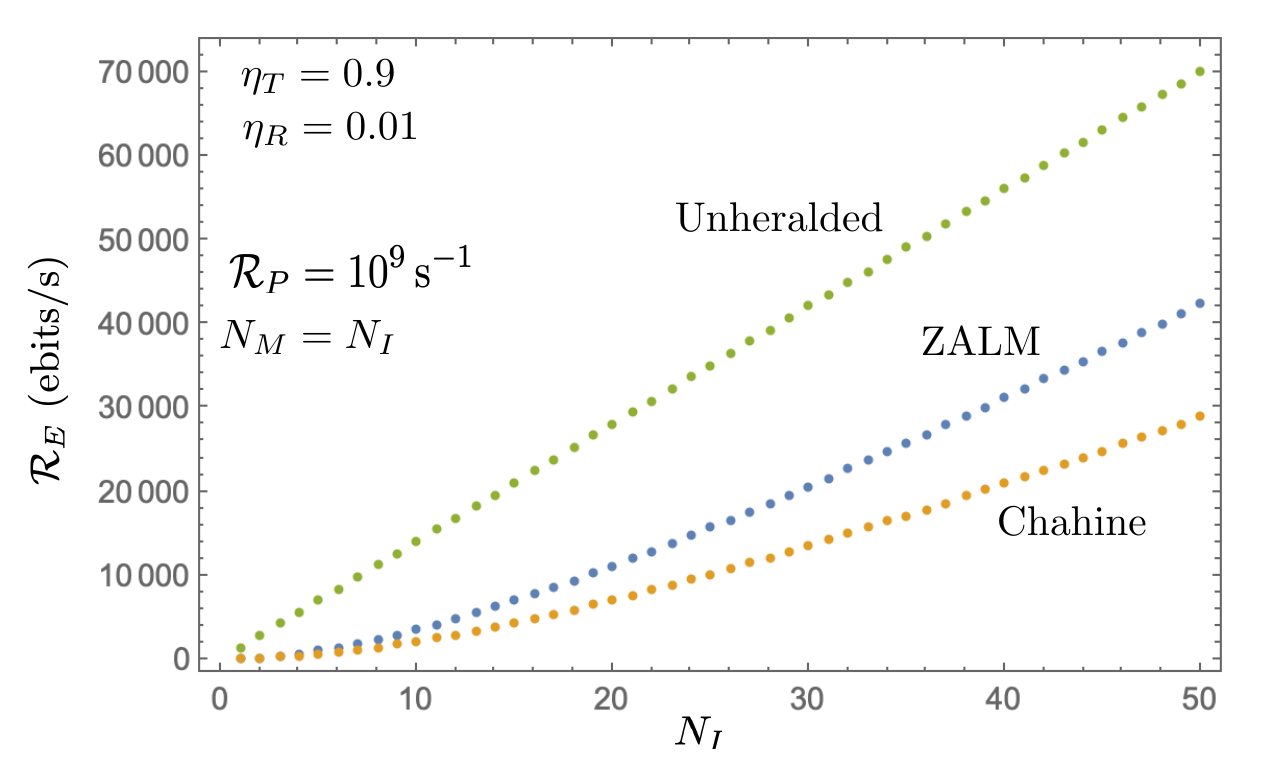}
\caption{Plots of the entanglement distribution rates, $\mathcal{R}_E$, in ebits/s, when Alice and Bob have $N_M=N_I$ memories allocated to each pump pulse, assuming $\mathcal{R}_P = 10^9\,{\rm s}^{-1}$, $\eta_T = 0.9$, and $\eta_R = 0.01$.  The $G-1$ values for all three configurations are given by the rightmost column in Table~\ref{dGvaluesNew}.  All three sources have $N_I$ phase-matched spectral islands. 
\label{ReNiMemories_fig}}    
\end{figure}

To complete our exploration of heralded versus unheralded operation we move on to third-generation implementations in which the numbers of source islands and QRX memories are unlimited.  Here $N_M = N_I$, and we get the distribution rates shown in Fig.~\ref{ReNiMemories_fig}.  This figure's $\mathcal{R}_E$ behavior differs from those shown in Figs.~\ref{ReSingleMemory_fig} and \ref{Re5Memories_fig}:  third-generation unheralded operation outperforms heralded operation, for $N_M = N_I$, with ZALM continuing to do better than Chahine~\emph{et al}.'s source.  Third-generation unheralded operation's superiority, despite $\Pr(|\psi^\pm\rangle_{\tilde{\bf S}}\mid \psi^\pm)$ being greater than $\Pr(|\psi^ -\rangle_{S_AI_B})$, is because $\mathbb{E}[\mathcal{H}(N_M)]/N_M$ is less than $\Pr(|\psi^ -\rangle_{S_AI_B})/\Pr(|\psi^\pm\rangle_{\tilde{\bf S}}\mid \psi^\pm)$ for the assumed parameter values.
    
\begin{figure}[hbt]
    \centering
\includegraphics[width=3.5in]{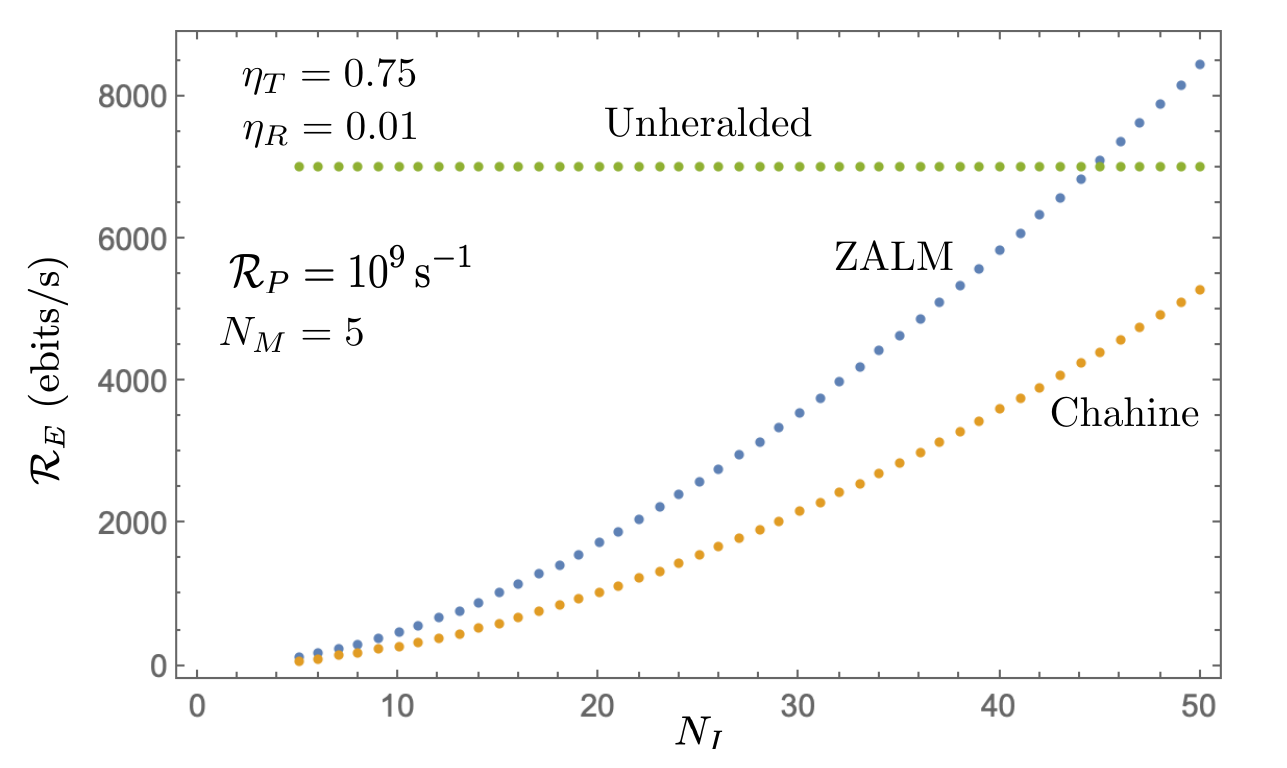}
\caption{Plots of the entanglement distribution rates, $\mathcal{R}_E$, in ebits/s, when Alice and Bob have 5 memories ($N_M=5$) allocated to each pump pulse, assuming $\mathcal{R}_P = 10^9\,{\rm s}^{-1}$, $\eta_T = 0.75$, and $\eta_R = 0.01$.  The $G-1$ values, $G-1 = 0.00686$ for ZALM, $G-1 = 0.0103$ for the Chahine~\emph{et al}.\@ source, and $G-1 = 0.00694$ for unheralded operation, ensure $\mathcal{B} \ge 0.999$ and $\mathcal{F} = 0.99$.   Unheralded operation's source has $N_M$ phase-matched spectral islands,  while the heralded sources have $N_I$ islands. 
\label{Re5Memories75_fig}}    
\end{figure}

For the cases we have considered in this section, we can say that:  (1) ZALM always outperforms the Chahine~\emph{et al.}\@ source; (2) first-generation ZALM ($N_I = 10$, $N_M=1$) provides a modest ($2.4$$\times$) distribution rate gain over single-island, single-memory unheralded operation; (3) second-generation ZALM ($N_I = 25$, $N_M = 5$) offers a similar ($2.2$$\times$) improvement over $N_I = N_M = 5$ unheralded operation; and (4)  third-generation unheralded operation ($N_I = N_M = 30$) enjoys a $2.0$$\times$ distribution rate advantage over ZALM.  That said, there are a few grains of salt that should be taken.  First, $\eta_T = 0.9$ is crucial to items (2) and (3); see Fig.~\ref{Re5Memories75_fig} where the $N_M = 5$ case is replotted with the QTX's efficiency reduced to $\eta_T = 0.75$, and the $G-1$ values for ZALM and the Chahine~\emph{et al}.\@ source changed to maintain $\mathcal{B} \ge 0.999$ and $\mathcal{F}=0.99$.  Now ZALM needs 50 spectral islands to eke out a $1.2$$\times$ distribution rate advantage over unheralded operation.  Remember that ZALM's $\eta_T< 1$ lumps together coupling loss from the ZALM's Sagnac sources to its partial-BSM apparatus with losses in the partial BSM's 50-50 beam splitter, its polarizing beam splitters, its DWDM filters, and its SPDs' subunit quantum efficiencies.  So 75\% overall efficiency may be more realistic than 90\% without aggressive optimization of all components and alignments.  On the other hand, if ZALM with $\eta_T = 0.9$ can be achieved, unheralded operation with $N_M=N_I = 100$ is only $1.4$$\times$ better than ZALM with $N_M = 5$ and $N_I = 100$.  We will continue discussion of ``to herald or not to herald'' in Sec.~\ref{Discussion} after we address some nonidealities that were ignored in Refs.~\cite{Shapiro2024,Shapiro2025} and, so far, here. 

\section{Dark Counts and Background Light \label{nonidealities}}
In this section we consider two nonidealities of islands-based entanglement distribution treated by neither Shapiro~\emph{et al}.~\cite{Shapiro2025} nor earlier in this paper, viz., dark counts in the heralded systems' SPDs, and background light collected in free-space propagation from the QTX to Alice and Bob's QRXs.  Dark counts change the heralding probabilities, $\Pr(\mathcal{H}_{H_n})$ and $\Pr(\mathcal{H}_{V_m})$, and, if those changes are significant, affect the joint state arriving at Alice and Bob's QRXs given there has been a herald.  Hence it can then impact all our performance metrics, i.e., $\mathcal{B},\mathcal{F}$, and $\mathcal{R}_E$.  Likewise, background light, if it is significant, will affect the joint state arriving at Alice and Bob's QRXs for all three of our configurations. Thus it too can impact all our performance metrics.   Our objectives here will be to show that: (1) dark counts will \emph{not} degrade the performance metrics reported in the previous section; and (2) background light can also be ignored \emph{if} the number of collected background modes is sufficiently low.

\subsection{Dark Counts in the Heralded Systems' SPDs}
Dark counts are usually taken to be Poisson distributed with rate $\mathcal{R}_D$, and statistically independent of photon detections.  The SPDs of choice for use in our heralded sources are SNSPDs, for which $\mathcal{R}_D =  10^3\,{\rm s}^{-1}$ or lower rates are commercially available.  Assuming that the detection interval is $T_D = 0.5\,$\,ns, which is both easily accomplished and compatible with the last section's assumption of $\mathcal{R}_P = 10^9{\rm s}^{-1}$, the average number of dark counts per SPD per pump pulse will be $\mathcal{R}_DT_D = 5\times 10^{-7}$.    Inasmuch as this number is orders or magnitude lower than the $G-1$ values from the rightmost column in Table~\ref{dGvaluesNew}, which were used in the last section's performance comparisons, we might immediately conclude that dark counts will not impact those performance results.  Here we will delve a little bit deeper in convincing ourselves that this is indeed true.

Assuming that the partial-PNR capable SPDs in the heralded systems' QTXs can distinguish between 0, 1, and $>$1 photon detections plus dark counts, the presence of dark counts changes the probability of an $n$th-island's declaration of an $H$-polarized detection and the probability of an $m$th-island's declaration of an $V$-polarized detection to
\begin{equation}
\Pr(\mathcal{H}_{H_n}) = \Pr(\mathcal{H}_{V_m}) = \frac{2\eta_T(G-1)e^{-2\mathcal{R}_DT_D}}{[\eta_T(G-1)+1]^3}
+ \frac{2\mathcal{R}_DT_De^{-2\mathcal{R}_DT_D}}{[\eta_T(G-1)+1]^2},
\end{equation}
for ZALM and
\begin{equation}
\Pr(\mathcal{H}_{H_n}) = \Pr(\mathcal{H}_{V_m})  = \frac{\eta_T(G-1)e^{-\mathcal{R}_DT_D}}{[\eta_T(G-1)+1]^2}
+ \frac{\mathcal{R}_DT_De^{-\mathcal{R}_DT_D}}{[\eta_T(G-1)+1]}, 
\end{equation}
for the Chahine~\emph{et al}.\@ source. The square of the first term in each expression is the probability of an $nm$th-island herald that is due to detection of one $n$th-island $H$-polarized photon and one $m$th-island $V$-polarized photon.  Half of that probability is then the probability of a \emph{true} $nm$th-island herald.  Table~\ref{darkcount_table} verifies that dark counts are inconsequential, insofar as these heralding probabilities are concerned, for the parameters used in Sec.~\ref{performanceC}.
\begin{table}
\centering
\begin{tabular}{|c||c|c|c|c|}\hline
 & \multicolumn{2}{c|}{$\Pr(\mathcal{H}_{H_n}) = \Pr(\mathcal{H}_{V_m})$} & \multicolumn{2}{c|}{$\Pr(\mathcal{H}_{nm}\mbox{ true})$}\\[.05in]\hline
 $\mathcal{R}_DT_D$ & ZALM & Chahine & ZALM & Chahine \\[.05in]\hline
 0 & 0.0297771&0.0225696 &0.000443337 &0.000254692 \\[.05in]
 $5 \times 10^{-7}$ & 0.0297780&0.0225770 &0.000443337 & 0.000254692\\[.05in]
 \hline
\end{tabular}
\caption{Probability of an $n$th-island $H$-polarized detection, $\Pr(\mathcal{H}_{H_n}) $, probability of an $m$th-island $V$-polarized detection, $\Pr(\mathcal{H}_{V_m})$, and the probability of a true $nm$th-island herald, $\Pr(\mathcal{H}_{nm}\mbox{ true})$, for ZALM and the Chahine~\emph{et al}.\@ source in the absence ($\mathcal{R}_DT_D = 0$) and presence ($\mathcal{R}_DT_D = 5\times 10^{-7}$) of dark counts.  The assumed detector efficiency is $\eta_T = 0.9$, and the $G-1$ values are taken from the rightmost column of Table~\ref{dGvaluesNew}. \label{darkcount_table}}
\end{table}

\subsection{Background Light Collected in Free-Space Propagation \label{Background}}
Background light is an important consideration for entanglement distribution from a satellite-borne QTX to ground station QRXs.  Daytime sky background at 1.55\,$\upmu$m wavelength has an average number, $N_B \sim 10^{-6}$, of photons per space-time-polarization mode~\cite{Shapiro2005}, with nighttime values being orders of magnitude lower.  Because $N_B \sim 10^{-6}$ is more than 2 orders of magnitude lower than the heralded sources' $\eta_R(G-1)$ values that give $\mathcal{B} \ge 0.999$ and $\mathcal{F} = 0.99$ when $\eta_T = 0.9$ and $\eta_R = 0.01$, we might conclude, as was the case for dark counts, that background light will not impact those sources' performance results from Sec.~\ref{performanceC}.  Similarly, because the unheralded source's $\eta_R(G-1)$ that gives  $\mathcal{B} \ge 0.999$ and $\mathcal{F} = 0.99$ when  $\eta_R = 0.01$ is more than 50$\times$ the nominal daytime $N_B$, we might expect that its performance would not suffer dramatically in the presence of background light.  Such is not the case.  Indeed \emph{all} of our configurations can suffer appreciable performance loss from background light, owing to background light's multimode nature.   A proper treatment of background light's effect on our configurations' entanglement distribution performance should include mode conversion and memory loading, topics that are beyond the scope of this paper.  Nevertheless, the analysis presented below will provide a preliminary indication how extraneous background modes, i.e., those other than the single temporal mode emitted by a source island, degrade our systems' performance.  

To include background light in our analysis we need to reexamine the beam splitter relationships used to account for QTX-to-QRX propagation losses in the absence of background, e.g., for the ZALM source's connection to Alice's QRX the relationship in question is
\begin{equation}
\hat{a}_{S_{A_P}} = \sqrt{\eta_R}\,\hat{a}_{S_{1_P}} + \sqrt{1-\eta_R}\,\hat{a}_{R_{1_P}}, 
\end{equation} 
for $P = H,V$.  In the absence of background light the $\hat{a}_{R_{1_P}}$ mode is in its 
vacuum state, but in background's presence that mode is in a thermal state with average photon number $N_B$.  Furthermore, although Alice and Bob's mode converters and quantum memories can reasonably be expected to respond only to a single \emph{spatial} mode, they may also admit an additional $M_B$ background \emph{temporal} modes per receiver per polarization that are in iid thermal states with average photon number $N_B$.     

In the presence of background light we continue to define a loadable state to be one for which both Alice and Bob's QRXs receive photons. Now, however, some or all of those photons may come from background light rather than QTX light.  As shown in Appendix~\ref{AppendC}, the probabilities of a loadable state are now:
\begin{equation}
\Pr({\rm loadable}) = 1-2N_S''^2\left[1-\frac{\eta_RN_S''}{2N_S}\right]^2\left[\frac{1}{N_B+1}\right]^{2M_B}  + N_S''^4\!\left[1-\frac{\eta_RN_S''}{N_S}\right]^2\left[\frac{1}{N_B+1}\right]^{4M_B},
\label{loadableZALMnb}
\end{equation}
for ZALM, where $N_S'' \equiv [\eta_R/N_S + (1-\eta_R)(N_B+1)]^{-1}$; 
\begin{equation}
\Pr({\rm loadable}) = 1-2\tilde{N}_B^{\prime 2}\!\left[1-\frac{\eta_R\tilde{N}'_B}{2N_S}\right]^2\!\left[\frac{1}{N_B+1}\right]^{2M_B} + 
N_S^{\prime\prime 2}\tilde{N}_B^2\!\left[1-\frac{\eta_RN''_S}{N_S}\right]^2\left[\frac{1}{N_B+1}\right]^{4M_B}, 
\label{loadableChahineNb}
\end{equation} 
for the Chahine~\emph{et al}.\@ source, where $\tilde{N}_B \equiv [1+(1-\eta_R)N_B]^{-1}$, and $\tilde{N}'_B \equiv 2N''_S\tilde{N}_B/(N''_S+\tilde{N}_B)$; and
\begin{align}
&\Pr({\rm loadable}) = 1-\frac{2}{[\eta_R G+(1-\eta_R)(N_B+1)]^2}\!\left[\frac{1}{N_B+1}\right]^{2M_B} \nonumber\\[.05in]
&+\frac{1}{2\eta_R(G-1-N_B)(N_B+1) + (N_B+1)^2 +\eta_R^2[(N_B+1)^2-G(2N_B+1)]^2}\!\left[\frac{1}{N_B+1}\right]^{4M_B},
\label{loadableUnheraldedNb}
\end{align}
for unheralded operation.  

\begin{figure}[hbt]
    \centering
\includegraphics[width=3.75in]{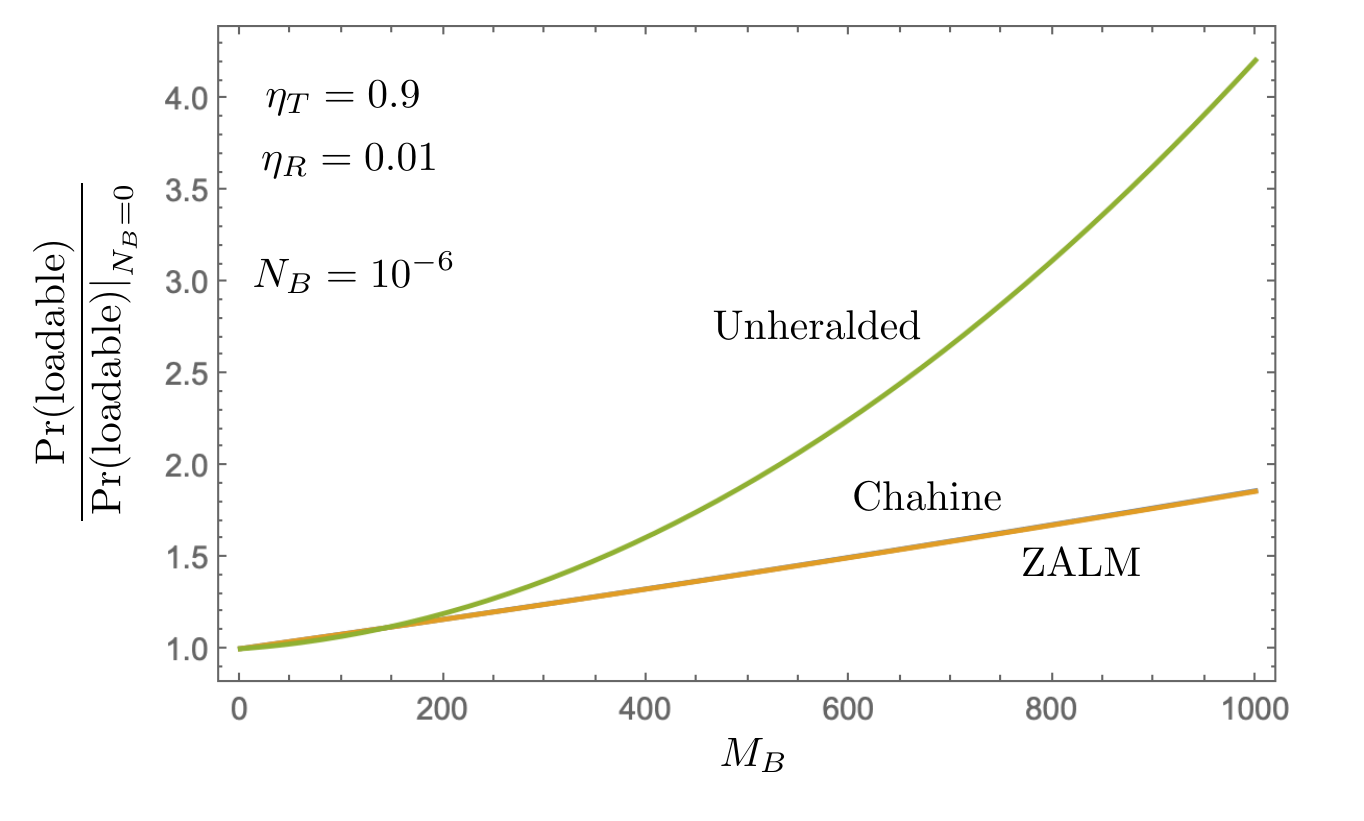}
\caption{Normalized plots of $\Pr({\rm loadable)}$ versus $M_B$, the number of extraneous modes per receiver per polarization, for ZALM, the Chahine~\emph{et al}.\@ source, and unheralded operation in the presence of background with average photon number $N_B = 10^{-6}$ per mode.  These curves assume $\eta_T = 0.9$, $\eta_R = 0.01$,  and the $G-1$ values from the rightmost column of Table~\ref{dGvaluesNew}.  The curves for ZALM and the Chahine~\emph{et al}.\@ source are  indistinguishable.  
\label{LoadableBackground_fig}}    
\end{figure}

Figure~\ref{LoadableBackground_fig}, which plots the loadable-state probabilities, normalized by their $N_B = 0$ values,  shows that $M_B \ge 100$ is needed for there to be an appreciable number of background-generated loadable states, and that for $M_B \gg 100$ unheralded operation becomes swamped by these extraneous loadable states.  Because $M_B = TW$, where $T$ is the background measurement's duration and $W$ is the background measurement's bandwidth, it is unlikely that $M_B \ge 100$ will occur.  For example, using $T=0.5\,$ns, the detection interval we assumed in our dark-count assessment, and $W = 50\,$GHz, a reasonable DWDM channel bandwidth, we get $M_B = 25$.  With this $M_B$ value, Fig.~\ref{LoadableBackground_fig}'s loadable-state probabilities for ZALM, the Chahine~\emph{et al}.\@ source, and unheralded operation are increased by 2.0\%, 2.0\%, and 1.2\%, respectively.  Small though these numbers may be, we still need to assess their impact on the Bell-state fraction and the Bell-state fidelity as we have set high goals for them, $\mathcal{B} \ge 0.999$ and $\mathcal{F} = 0.99$.

Appendix~\ref{AppendC} shows that Eqs.~(\ref{PrCorrectZALM})--(\ref{BellZALM}) multiplied by $[1/(N_B+1)]^{4M_B}$ and with $N_S'$ replaced by $N_S''$ give the Bell-state probabilities for ZALM in the presence of background light.  Greater changes are needed to include background light in the Bell-state probabilities for the Chahine~\emph{et al}.\@ source and unheralded operation.  For these configurations Appendix~\ref{AppendC} gives:
\begin{align}
\Pr(|\psi^-\rangle_{\tilde{\bf S}}\mid \psi^-) &= N_S^{\prime\prime 2}\tilde{N}_B^2\!\left[ \frac{(1- N_B')^2 +(1-N_S'')^2}{2} -\frac{\eta_RN_S''[(1-N_B')^2 + (1-2N_S'')(1-N_S'')]}{N_S} \right.\nonumber \\[.05in]
&\,\,+ \left.\frac{\eta_R^2N_S^{\prime\prime 2}[(1-N_B')^2+(1-2N_S'')^2]}{2N_S^2}
\right]\!\left[\frac{1}{N_B+1}\right]^{4M_B},
\label{PrCorrectChahineNb}
\end{align}
\begin{align}
\Pr&(|\psi^+\rangle_{\tilde{\bf S}}\mid \psi^-) = \nonumber \\[.05in]
&\,\, N_S^{\prime\prime 2}N_B'^2(1-N_B')\left[(1-N_S'') - \frac{\eta_RN_S''(2-3N_S'')}{N_S} +\frac{\eta_R^2N_S^{\prime\prime 2}(1-2N_S'')}{N_S^2}\right]\!\left[\frac{1}{N_B+1}\right]^{4M_B}, 
\end{align}
and
\begin{align}
\Pr(|\phi^\pm\rangle_{\tilde{\bf S}}\mid \psi^-) &= N_S^{\prime\prime 2}\tilde{N}_B^2\!\left[\frac{(1-N_B')^2+(1-N_S'')^2}{2} -\frac{\eta_RN_S''[(1-N_B')^2+(1-2N_S'')(1-N_S'')]}{N_S} \right. \nonumber \\[.05in]
& \,\, + \left.\frac{\eta_R^2N_S^{\prime\prime 2}[(1-N_B')+(1-3N_S'')(1-N_S'')}{2N_S^2}\right]\!\left[\frac{1}{N_B+1}\right]^{4M_B},
\label{PrErrorChahineNb}
\end{align}
for the Chahine~\emph{et al}.\@ source, where $N_B'\equiv 1/(N_B+1)$; 
and
\begin{equation}
\Pr(|\psi^-\rangle_{\tilde{\bf S}}\mid \psi^-) = \frac{(1-2D_1)^{\prime 2} + 8 D_2^{\prime 2}}{D_3^{\prime 2}(N_B+1)^{4M_B}},
\label{PrCorrectUnheraldedNb}
\end{equation}
\begin{equation}
\Pr(|\psi^+\rangle_{\tilde{\bf S}}\mid \psi^-) = \Pr(|\phi^+\rangle_{\tilde{\bf S}}\mid \psi^-) = 
\Pr(|\phi^-\rangle_{\tilde{\bf S}}\mid \psi^-) = \frac{(1-2D_1)^{\prime 2}}{D_3^{\prime 2}(N_B+1)^{4M_B}},
\label{PrErrorUnheraldedNb}
\end{equation}
for unheralded operation, where
\begin{equation}
D_1' \equiv [1+\eta_R(G-1-N_B)]/2D_3',
\end{equation}
\begin{equation}
D_2' \equiv \eta_R\sqrt{G(G-1)}/2D_3',
\end{equation}
and
\begin{equation}
D_3'\equiv 2\eta_R(G-1-N_B)(N_B+1)+(N_B+1)^2+ \eta_R^2[(N_B+1)^2-G(2N_B+1)].
\end{equation}  

\begin{figure}[hbt]
    \centering
\includegraphics[width=3.75in]{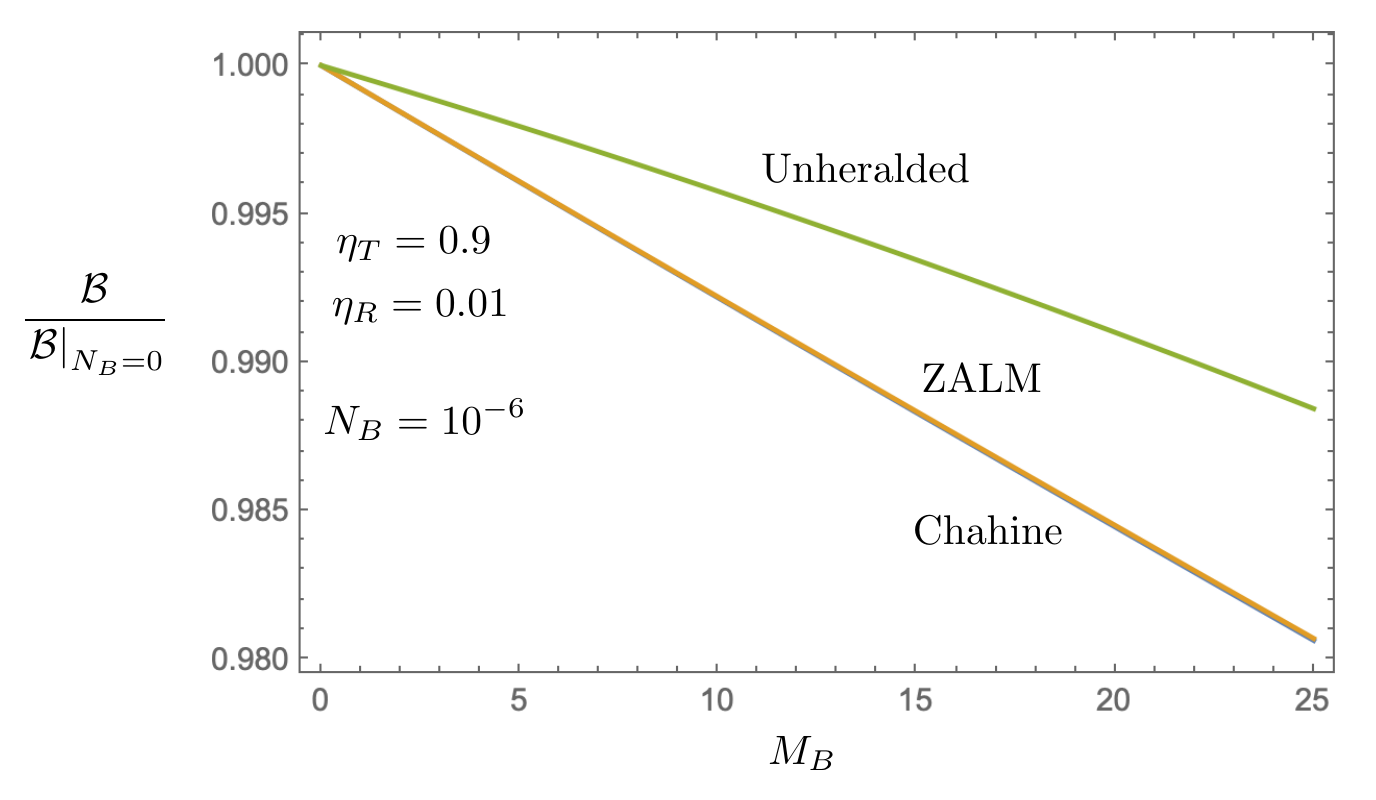}
\caption{Normalized plots of $\mathcal{B}$ versus $M_B$, the number of extraneous modes per receiver per polarization, for ZALM, the Chahine~\emph{et al}.\@ source, and unheralded operation in the presence of background with average photon number $N_B = 10^{-6}$ per mode.  These curves assume $\eta_T = 0.9$, $\eta_R = 0.01$,  and the $G-1$ values from the rightmost column of Table~\ref{dGvaluesNew}.  The curves for ZALM and the Chahine~\emph{et al}.\@ source are indistinguishable.
\label{BellFractionBackground_fig}}    
\end{figure}

Figure~\ref{BellFractionBackground_fig} plots, versus $M_B$, the Bell-state fractions,  normalized by their $N_B = 0$ values, for ZALM, the Chahine~\emph{et al}.\@ source, and unheralded operation in the presence of background with average photon number $N_B = 10^{-6}$ per mode.  These curves assume $\eta_T = 0.9$, $\eta_R = 0.01$, and the $G-1$ values from the rightmost column of Table~\ref{dGvaluesNew}.  They show that ZALM and the Chahine~\emph{et al}.\@ source retain $>$99\% Bell-state fraction up to $M_B = 12$, and unheralded operation does the same up to $M_B = 21$. Thus all three configurations' Bell-state fractions have significant robustness to background light, e.g., our $T=0.5\,$ns, $W= 50\,$GHz example has $M_B = 25$, so reducing $T$ and $W$ to the still reasonable values 0.25\,ns and 25\,GHz drops $M_B$ to 6.25.    

Our configurations' Bell-state fidelities in the presence of background light do not depend on $M_B$ as only light received in the QTX modes affect $\mathcal{F}$. Table~\ref{BellStateFidelityNb} shows the fidelity degradations that ensue, for our standard ($\eta_T = 0.9$, $\eta_R = 0.01$, and $G-1$ values taken from the rightmost column of Table~\ref{dGvaluesNew}) example, when $N_B$ goes from 0 to $10^{-6}$.
\begin{table}
\centering
\begin{tabular}{|c||c|c|}\hline
Configuration & $N_B = 0$ & $N_B = 10^{-6}$ \\[.05in] \hline
ZALM &0.99 &0.9894 \\[.05in]
Chahine &0.99 &0.9892 \\[.05in]
Unheralded &0.99 &0.9897\\[.05in] \hline
\end{tabular}
\caption{Bell-state fidelities for ZALM, the Chahine~\emph{et al}.\@ source, and unheralded operation in the absence ($N_B=0$) and presence ($N_B = 10^{-6}$) of background light.  These fidelities assume $\eta_T = 0.9$, $\eta_R = 0.01$. and $G-1$ values taken from the rightmost column of Table~\ref{dGvaluesNew}. \label{BellStateFidelityNb}}
\end{table}

The upshot of our background-light analysis is that all three of our configurations have promising robustness for daytime operation at 1.55\,$\upmu$m wavelength if they can restrict the number of extraneous modes they collect to $\sim$10 or less.  

\section{Conclusions and Discussion \label{Discussion}}
Our paper's goal, addressing the ``to herald or not to herald'' issue for entanglement distribution using islands-based SPDCs, has been largely accomplished.  Admittedly, we did not make a comprehensive exploration of the parameter space, but we have developed the theory in sufficient detail to permit such work to be accomplished.  Moreover, the examples we have presented reveal behaviors that, ignoring disparities (see below) in the implementation cost or complexity for ZALM, the Chahine~\emph{et al}.\@ source, or unheralded operation, support the following conclusions.  First, ZALM's entanglement distribution rate dominates that of Chahine~\emph{et al}.'s source.  Second, when Alice and Bob allocate only a single memory per pump pulse, ZALM's entanglement distribution rate at high partial-BSM efficiency exceeds that of unheralded operation. Third, when memories are unlimited,  unheralded operation provides the highest entanglement distribution rate.  Now let us pay brief attention to some implementation issues.

The foundation for this paper was laid in Shapiro~\emph{et al}.~\cite{Shapiro2025}, where islands-based SPDCs were assumed to have identical, except for center frequency, factorable phase-matching islands that were pumped to emit the same average number of signal-idler photon pairs.  This behavior is no mean feat.  Any departure therefrom will introduce some distinguishability for the SPCI-heralded configurations.  Thus the Chahine~\emph{et al}.\@ source will enjoy an advantage over ZALM in spectrally-matching its islands because its single Sagnac source's cw and ccw SPDCs employ the same nonlinear crystal, whereas ZALM must match the islands of the cw and ccw SPDCs from two nonlinear crystals.   Even with perfect island matching, both systems' SPCI heralding incurs some loss of indistinguishability when the $G-1$ values for all islands are not identical.  Interestingly, neither of these issues significantly impact multi-memory unheralded operation, because it uses a single Sagnac source, and $G-1$ variations across different islands only detract from $\mathcal{R}_E$'s being proportional to $N_M$.

\begin{figure}[hbt]
    \centering
\includegraphics[width=4.5in]{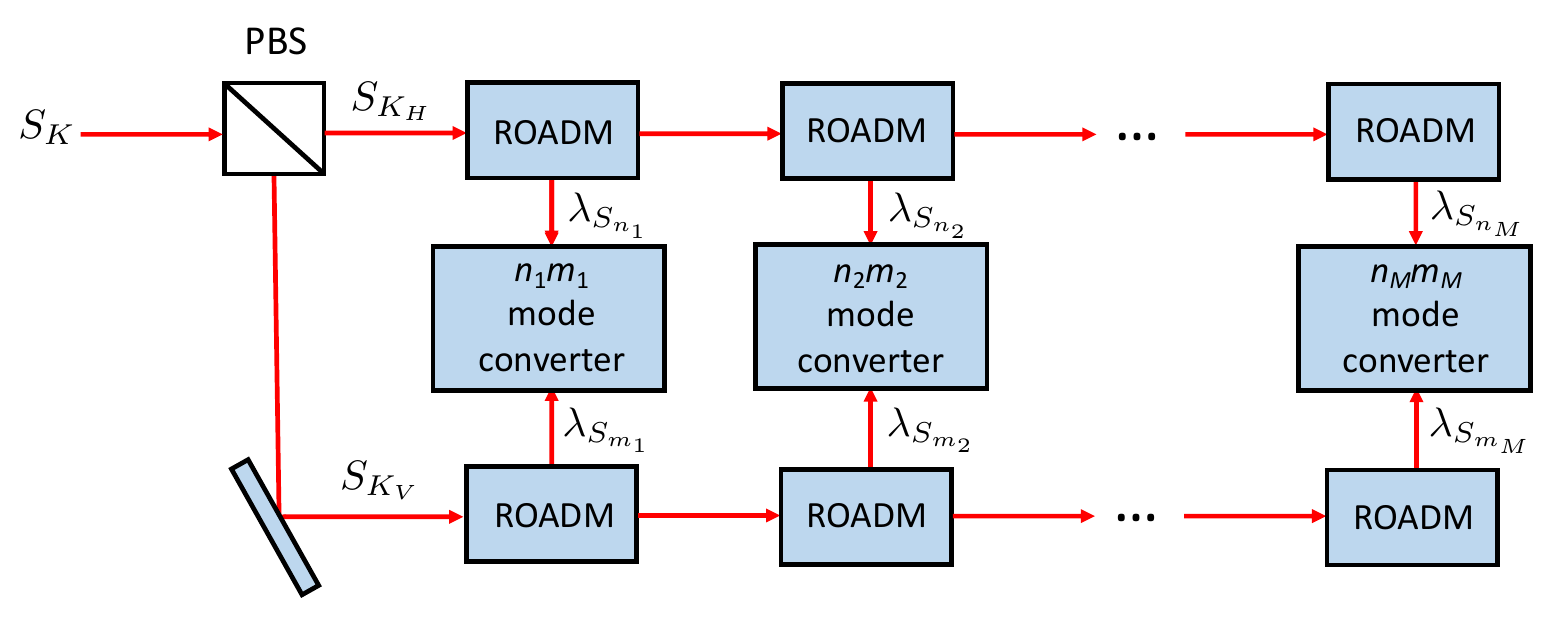}
\caption{Alternative architecture for $nm$th-island mode conversion for heralded systems whose QRXs have $N_M >1$ memories and whose QTX's SPDCs have $N_I \gg N_M$ phase-matched spectral islands.  The $n_km_k$th mode converter first does $H$-polarized wavelength conversion from $\lambda_{S_{n_k}}$ to $\lambda_M$ and  $V$-polarized  wavelength conversion from $\lambda_{S_{m_k}}$ to $\lambda_M$.  It then combines the outputs using a PBS and does temporal-mode shaping on the combined beam.  The final, mode-shaped output is then directed to the $k$th quantum memory.  $S_K$:  signal light arriving at Alice ($K=A$) and Bob's ($K=B$) QRXs.  $\lambda_M$:  quantum memory wavelength.  PBS:  polarizing beam splitter.  ROADM:  reconfigurable optical add-drop multiplexer.
\label{ROADM_fig}}    
\end{figure}

Mode conversion, both wavelength and temporal, is another area in which implementation issues are worthy of consideration.  For single-memory operation ZALM, Chahine~\emph{et al}., and unheralded operation all require only a single mode-converter, but ZALM and Chahine~\emph{et al}.'s must be rapidly tunable across the full span of their islands' center wavelengths, whereas unheralded operation's mode converter does not require tunability.  Multi-memory unheralded operation requires a dedicated mode converter for each memory, and, as stated in Sec.~\ref{performanceD}, ZALM with $N_I = 100$ and $N_M = 5$ approaches unheralded operation's $N_I=N_M =100$ performance with 1/20th the memory resources.  Of course, this assessment ignores the path-switching losses in multi-memory heralded systems' mode converters.  Thus, the alternative to the Fig.~\ref{mode-conversion1_fig} architecture for multi-memory heralded reception shown in Fig.~\ref{ROADM_fig} may be of interest.  It uses two chains of reconfigurable optical add-drop multiplexers (ROADMs), one for the $H$-polarized arriving light and the other for the $V$-polarized light, to sequentially drop the light associated with the $n_km_k$th-island herald from the chain for wavelength conversion and temporal mode shaping.  The resulting light after those operations is then directed to the $k$th quantum memory for loading.  Unfortunately, commercially-available ROADMs currently have $\sim$10\,dB insertion loss.

Our final remarks are devoted to topics for follow-on work.  The first is a more comprehensive performance evaluation---for ZALM and unheralded operation, because the Chahine~\emph{et al}.\@ source is dominated by one or the other---that is linked to realistic parameters for domain-engineered islands-based SPDCs and the alternative intra-cavity SPDCs using standard periodically-poled crystals.  Domain engineering has already demonstrated an 8-island downconverter~\cite{Morrison2022}, but it is not clear how much further that technology can be pushed.  Moreover, their islands' multi-GHz bandwidths place a heavy burden on the bandwidth compression needed to mode match their outputs to the multi-MHz bandwidth of color-center quantum memories.  In contrast intra-cavity SPDCs can greatly reduce the required bandwidth compression~\cite{Rambach2016}, and perhaps even obviate its use, but their internal losses and device complexity may be problematic.  

A second area worthy of follow-on work is extending the results presented herein to include Duan-Kimble and emissive memory loading for both heralded and unheralded operation.  Shapiro~\emph{et al}.'s treatment of Duan-Kimble loading~\cite{Shapiro2024,Raymer2024}, which assumes ideal mode conversion and unit Bell-state fraction, should be directly applicable to the islands-based sources we have treated here.  Emissive loading, however, will require a new analysis.  Comparing the entanglement distribution rates for these two loading procedures will be valuable in that Duan-Kimble loading requires a difficult-to-fabricate intra-cavity quantum memory, but, in the absence of dark counts, all click patterns in which both Alice and Bob make detections herald memory loading.  Emissive loading's quantum memories are easier to fabricate, but both Alice and Bob must have successful partial BSMs to herald loading.  Thus, in lossless operation with ideal mode conversion and no dark counts, emissive loading suffers a factor-of-1/4 reduction in the entanglement distribution rate from what we have reported here, whereas Duan-Kimble loading would not under similarly ideal conditions.

\appendix

\section{Derivations for the Chahine~\emph{et al}.\@ Source \label{AppendA}}
This appendix provides derivation details for Eqs.~(\ref{PrLoadableChahine})--(\ref{PrErrorChahine}).  Starting from Eq.~(\ref{antinorm_Chahine}), we have that
\begin{equation}
\chi^{\rho_{SS^\perp}\mid \psi^-}_A(\bzeta) = \chi_A^{\rho_{S\mid\psi^-}}(\bzeta_S) 
\chi^{\rho_{S^\perp}}_A(\bzeta_{S^\perp}) 
= e^{-\bzeta_S^\dagger\bzeta_S/N_S}\left[1-\frac{|\zeta_{S_H}|^2}{N_S}\right]\left[1-\frac{|\zeta_{S_V}|^2}{N_S}\right]e^{-\bzeta^\dagger_{S^\perp}\bzeta_{S^\perp}},
\end{equation}
where $\bzeta^\dagger \equiv [\begin{array}{cc}\bzeta^\dagger_S & \bzeta^\dagger_{S^\perp}\end{array}]$, and $\bzeta^\dagger_K \equiv [\begin{array}{cc} \zeta^*_{K_H} & \zeta^*_{K_V}\end{array}]$ for $K = S, S^\perp$, is the anti-normally ordered characteristic function for the light entering the Chahine~\emph{et al}.\@ source's signal-path erasing beam splitter given there has been a $\psi^-$ herald.  Applying the inverse of that beam splitter's input-output relations, i.e.,
\begin{equation}
\hat{a}_{S_P} = \frac{\hat{a}_{S_{1_P}} + \hat{a}_{S_{2_P}}}{\sqrt{2}} \quad\mbox{and}\quad
\hat{a}_{S^\perp_P} = \frac{\hat{a}_{S_{1_P}} - \hat{a}_{S_{2_P}}}{\sqrt{2}},
\mbox{ for $P = H,V$},
\end{equation}
we have that 
\begin{align}
\chi^{\rho_{{\bf S}'} \mid \psi^-}_A(\bzeta) &= 
e^{-(|\zeta_{S_{1_H}}+\zeta_{S_{2_H}}|^2+|\zeta_{S_{1_V}}+\zeta_{S_{2_V}}|^2)/2N_S}
e^{-(|\zeta_{S_{1_H}}-\zeta_{S_{2_H}}|^2+|\zeta_{S_{1_V}}-\zeta_{S_{2_V}}|^2)/2} \nonumber \\[.05in]
& \,\,\times \left[1-\frac{|\zeta_{S_{1_H}}+\zeta_{S_{2_H}}|^2}{2N_S}\right]\left[1-\frac{|\zeta_{S_{1_V}}+\zeta_{S_{2_V}}|^2}{2N_S}\right],
\end{align}
is the anti-normally ordered characteristic function for the ${\bf S}' \equiv S_1S_2$ modes sent to Alice and Bob.  After propagation to Alice and Bob's QRXs, the preceding characteristic function yields
\begin{align}
\chi^{\rho_{\tilde{\bf S}} \mid \psi^-}_A(\bzeta) &= 
e^{-(|\zeta_{S_{A_H}}+\zeta_{S_{B_H}}|^2+|\zeta_{S_{A_V}}+\zeta_{S_{B_V}}|^2)/2N_S'}
e^{-(|\zeta_{S_{B_H}}-\zeta_{S_{B_H}}|^2+|\zeta_{S_{A_V}}-\zeta_{S_{B_V}}|^2)/2} \nonumber \\[.05in]
& \,\,\times \left[1-\frac{\eta_R|\zeta_{S_{A_H}}+\zeta_{S_{B_H}}|^2}{2N_S}\right]\left[1-\frac{\eta_R|\zeta_{S_{A_V}}+\zeta_{S_{B_V}}|^2}{2N_S}\right],
\label{ChahineAB}
\end{align}
for their $\tilde{\bf S} \equiv S_AS_B$ modes.

It is a straightforward matter to derive Eq.~(\ref{PrLoadableChahine}) from Eq.~(\ref{ChahineAB}).  By its definition we have
\begin{equation}
\Pr({\rm loadable}) = 1 - {\rm Tr}_{S_B}[{}_{S_A}\langle {\bf 0}|\hat{\rho}_{\tilde{\bf S}\mid \psi^-}|{\bf 0}\rangle_{S_A}] -  {\rm Tr}_{S_A}[{}_{S_B}\langle {\bf 0}|\hat{\rho}_{\tilde{\bf S}\mid \psi^-}|{\bf 0}\rangle_{S_B}] + {}_{\tilde{\bf S}}\langle {\bf 0}|\hat{\rho}_{\tilde{\bf S}\mid \psi^-}|{\bf 0}\rangle_{\tilde{\bf S}}.
\end{equation}
Doing the vacuum-state projections on the 8D operator-valued inverse Fourier transform of Eq.~(\ref{ChahineAB}), and then taking the traces in the coherent-state basis, results in 
\begin{equation}
{\rm Tr}_{S_B}[{}_{S_A}\langle {\bf 0}|\hat{\rho}_{\tilde{\bf S}\mid \psi^-}|{\bf 0}\rangle_{S_A}] =   {\rm Tr}_{S_A}[{}_{S_B}\langle {\bf 0}|\hat{\rho}_{\tilde{\bf S}\mid \psi^-}|{\bf 0}\rangle_{S_B}] = \tilde{N}_S^2\!\left(1-\frac{\eta_R\tilde{N}_S}{2N_S}\right)^2,
\end{equation}
which, together with
\begin{equation}
{}_{\tilde{\bf S}}\langle {\bf 0}|\hat{\rho}_{\tilde{\bf S}\mid \psi^-}|{\bf 0}\rangle_{\tilde{\bf S}} = N_S^{\prime 2}\!\left(1-\frac{\eta_RN'_S}{N_S}\right)^2,
\end{equation}
gives us Eq.~(\ref{PrLoadableChahine}).

The direct route to obtaining Eqs.~(\ref{PrCorrectChahine})--(\ref{PrErrorChahine}) from Eq.~(\ref{ChahineAB}) is to project the latter's 8D operator-valued inverse Fourier transform onto the ${\tilde{\bf S}}$ Bell states.  We, however, find it more convenient to recognize that the propagation beam splitters commute with the signal-path erasing beam splitter.  Thus we will project
\begin{equation}
\hat{\rho}_{S'S^{\prime\perp}\mid\psi^-} \equiv \int\!\frac{{\rm d}^8\bzeta}{\pi^4}\,
e^{-\bzeta_{S'}^\dagger\bzeta_{S'}/N_S'}e^{-\bzeta^\dagger_{S^{\prime\perp}}\bzeta_{S^{\prime\perp}}}\left[1-\frac{\eta_R|\zeta_{S'_H}|^2}{N_S}\right]\left[1-\frac{\eta_R|\zeta_{S'_V}|^2}{N_S}\right]e^{-\bhata^\dagger\bzeta}e^{\bzeta^\dagger\bhata},
\label{backpropagatedChahineNoNb}
\end{equation}
onto the beam-splitter transformed versions of the $S_AS_B$ Bell states.  For example, writing 
\begin{equation}
|\psi^-\rangle_{\tilde{\bf S}} = (\hat{a}_{A_H}^\dagger\hat{a}_{B_V}^\dagger -\hat{a}_{A_V}^\dagger\hat{a}_{B_H}^\dagger)|{\bf 0}\rangle_{\tilde{\bf S}}/\sqrt{2}, 
\label{singletdefn}
\end{equation}
the transformed state onto which we project $\hat{\rho}_{S'S^{\prime\perp}\mid\psi^-}$ is obtained by using
\begin{equation}
\hat{a}_{S_{A_P}}^\dagger = \frac{\hat{a}_{S'_P}^\dagger + \hat{a}_{S^\perp_P}^\dagger}{\sqrt{2}},\mbox{ for $P=H,V$,}
\end{equation}
and
\begin{equation}
\hat{a}_{S_{B_P}}^\dagger = \frac{\hat{a}_{S'_P}^\dagger - \hat{a}_{S^\perp_P}^\dagger}{\sqrt{2}},\mbox{ for $P=H,V$,}
\end{equation}
in Eq.~(\ref{singletdefn}).  

After those projections have been accomplished we are left with each Bell-state probability being expressed as an 8D integral of an 8th-order polynomial in $\{\zeta_{S'_P},\zeta_{S^{\prime\perp}_P}: P = H,V\}$ multiplied by 
$N_S^{\prime 2}[
e^{-\bzeta_{S'}^\dagger\bzeta_{S'}/N_S'}/(\pi N_S')^2][e^{-\bzeta^\dagger_{S^{\prime\perp}}\bzeta_{S^{\prime\perp}}}/\pi^2]$.  These integrals, evaluated by a tedious complex-Gaussian moment factoring process, give us Eqs.~(\ref{PrCorrectChahine})--(\ref{PrErrorChahine}).  For additional details on the moment factoring leading to Eqs.~(\ref{PrCorrectChahine})--(\ref{PrErrorChahine}), and the similar moment-factoring calculations needed to show that the off-diagonal elements in the Bell-state representation of $\hat{\rho}^{(\mathcal{H_B})}_{\tilde{S}\mid \psi^-}$ all vanish, see Ref.~\cite{data}.

\section{Derivations for Unheralded Operation \label{AppendB}}
In this appendix we derive Eqs.~(\ref{PrLoadableUnheralded})--(\ref{PrErrorUnheralded}) using an approach paralleling what was done in Ref.~\cite{Shapiro2025} for ZALM.  First, we rewrite Eq.~(\ref{unheralded_chiAloss}) as
\begin{equation}
\chi_A^{\rho_{S_AI_B}}(\bzeta) = \exp[-\bzeta^{(R)T}\tilde{\Lambda}^{(R)}_{S_AI_B}\bzeta^{(R)}/2]\exp[-\bzeta^{(I)T}\tilde{\Lambda}^{(I)}_{S_AI_B}\bzeta^{(I)}/2],
\label{chiAunheraldedRI}
\end{equation}
where $\bzeta^{(R)}$ and $\bzeta^{(I)}$ are the real and imaginary parts of $\bzeta$, 
\begin{equation}
\tilde{\Lambda}^{(R)}_{S_AI_B} = \left[\begin{array}{cccc} 2[\eta_R(G-1)+1] & 0 & 0 & 2\eta_R\sqrt{G(G-1)} \\[.05in]
0 & 2[\eta_R(G-1)+1] & -2\eta_R\sqrt{G(G-1)} & 0 \\[.05in]
0 & -2\eta_R\sqrt{G(G-1)} & 2[\eta_R(G-1)+1] & 0 \\[.05in]
2\eta_R\sqrt{G(G-1)} & 0 & 0 & 2[\eta_R(G-1) + 1]
\end{array}\right],
\label{tildeLamRnoNb}
\end{equation}
and
\begin{equation}
\tilde{\Lambda}^{(I)}_{S_AI_B} = \left[\begin{array}{cccc} 2[\eta_R(G-1)+1] & 0 & 0 & -2\eta_R\sqrt{G(G-1)} \\[.05in]
0 & 2[\eta_R(G-1)+1] & 2\eta_R\sqrt{G(G-1)} & 0 \\[.05in]
0 & 2\eta_R\sqrt{G(G-1)} & 2[\eta_R(G-1)+1] & 0 \\[.05in]
-2\eta_R\sqrt{G(G-1)} & 0 & 0 & 2[\eta_R(G-1) + 1] \end{array}\right].
\label{tildeLamInoNb}
\end{equation}
We can now find $\Pr({\rm loadable})$ by starting from 
\begin{equation}
\Pr({\rm loadable}) = 1 - {\rm Tr}_{I_B}[{}_{S_A}\langle {\bf 0}|\hat{\rho}_{S_AI_B}|{\bf 0}\rangle_{S_A}] - {\rm Tr}_{S_A}[{}_{I_B}\langle {\bf 0}|\hat{\rho}_{S_AI_B}|{\bf 0}\rangle_{I_B} ] + {}_{S_AI_B}\langle {\bf 0}|\hat{\rho}_{S_AI_B}|{\bf 0}\rangle_{S_AI_B}.
\end{equation}
As done in Appendix~\ref{AppendA}, we first do the vacuum-state projections on the 8D operator-valued inverse Fourier transform of Eq.~(\ref{chiAunheraldedRI}), and then take the traces in the coherent-state basis, to get
\begin{equation}
{\rm Tr}_{I_B}[{}_{S_A}\langle {\bf 0}|\hat{\rho}_{S_AI_B}|{\bf 0}\rangle_{S_A}] = {\rm Tr}_{S_A}[{}_{I_B}\langle {\bf 0}|\hat{\rho}_{S_AI_B}|{\bf 0}\rangle_{I_B} ] =
\frac{1}{[\eta_R(G-1)+1]^2},
\end{equation}
and
\begin{equation}
{}_{S_AI_B}\langle {\bf 0}|\hat{\rho}_{S_AI_B}|{\bf 0}\rangle_{S_AI_B} = 
\frac{1}{[\eta_R(2-\eta_R)(G-1)+1]^2},
\end{equation}
from which Eq.~(\ref{PrLoadableUnheralded}) follows.

To find unheralded operation's Bell-state probabilities, we enable a Gaussian moment-factoring approach by rewriting $\chi_A^{\rho_{S_AI_B}}(\bzeta)$ as
\begin{equation}
\chi_A^{\rho_{S_AI_B}}(\bzeta) = \exp[-\bzeta^{(R)T}(\Lambda^{(R)}_{S_AI_B})^{-1}\bzeta^{(R)}/2]\exp[-\bzeta^{(I)T}(\Lambda^{(I)}_{S_AI_B})^{-1}\bzeta^{(I)}/2],
\end{equation}
where
\begin{equation}
\Lambda^{(R)}_{S_AI_B} = \left[\begin{array}{cccc} D_1 & 0 & 0 & -D_2 \\[.05in]
0 & D_1 & D_2 & 0 \\[.05in]
0 & D_2 & D_1 & 0 \\[.05in]
-D_2 & 0 & 0 & D_1 \end{array}\right],
\end{equation}
and
\begin{equation}
\Lambda^{(I)}_{S_AI_B} = \left[\begin{array}{cccc} D_1 & 0 & 0 & D_2 \\[.05in]
0 & D_1 & -D_2 & 0 \\[.05in]
0 & -D_2 & D_1 & 0 \\[.05in]
D_2 & 0 & 0 & D_1 \end{array}\right],
\end{equation}
with
\begin{equation}
D_1 \equiv [\eta_R(G-1)+1]/2D_3,
\label{D1defn}
\end{equation}
\begin{equation}
D_2 \equiv \eta_R\sqrt{G(G-1)}/2D_3,
\label{D2defn}
\end{equation}
\begin{equation}
D_3 \equiv 1-\eta_R(\eta_R-2)(G-1),
\label{D3defn}
\end{equation}
and
\begin{equation}
{\rm Det}[\Lambda^{(R)}_{S_AI_B}] = {\rm Det}[\Lambda^{(I)}_{S_AI_B}] = \frac{1}{16 D_3^2}.
\end{equation}
Now, $\hat{\rho}_{S_AI_B}$ can be expressed as
\begin{equation}
\hat{\rho}_{S_AI_B} = 16\sqrt{{\rm Det}[\Lambda^{(R)}_{S_AI_B}]{\rm Det}[\Lambda^{(I)}_{S_AI_B}]}\int\!{\rm d}^8\bzeta\,p_R(\bzeta^{(R)})\,p_I(\bzeta^{(I)})\,e^{-\bhata^\dagger\bzeta}e^{\bzeta^\dagger\bhata},
\label{rhoSaIbUnheralded}
\end{equation}
where $p_K(\bzeta^{(K)})$ for $K = R,I$ is the joint probability density function for 4 real-valued, zero-mean, (fictitious) jointly-Gaussian random variables with covariance matrix $\Lambda^{(K)}_{S_AI_B}$.  Bell-state projections of $e^{-\bhata^\dagger\bzeta}e^{\bzeta^\dagger\bhata}$ are 4D polynomials of $\{\bzeta_{S_A},\bzeta_{I_B}\}$, resulting in integrals for unheralded operation's Bell-state probabilities that can be evaluated from Gaussian moment factoring. The answers so obtained are 
\begin{equation}
\Pr(|\psi^-\rangle_{\tilde{\bf S}}\mid \psi^-) = [(1-2D_1)^2 + 8 D_2^2]/D_3^2,
\label{PrCorrectUnheraldedRaw}
\end{equation}
and
\begin{equation}
\Pr(|\psi^+\rangle_{\tilde{\bf S}}\mid \psi^-) = \Pr(|\phi^+\rangle_{\tilde{\bf S}}\mid \psi^-) = 
\Pr(|\phi^-\rangle_{\tilde{\bf S}}\mid \psi^-) = (1-2D_1)^ 2/D_3^2,
\label{PrErrorUnheraldedRaw}
\end{equation}
Plugging Eqs.~(\ref{D1defn})--(\ref{D3defn}) into these probabilities, and doing some algebraic simplifications, verifies Eqs.~(\ref{PrCorrectUnheralded}) and (\ref{PrErrorUnheralded}).  Similar calculations show that the off-diagonal elements in the Bell-state representation of $\hat{\rho}^{(\mathcal{H}_B)}_{S_AI_B}$ vanish.  For more details about this moment factoring see Ref.~\cite{data}.

\section{Derivations for Background Light Effects \label{AppendC}}
This paper's last task is to supply some derivation details for Sec.~\ref{Background}'s results.  The results for ZALM are trivially obtained.  For ZALM the anti-normally ordered characteristic function for the signal modes received by Alice and Bob, given there has been a herald, is
\begin{equation}
\chi_A^{\rho_{\tilde{\bf S}\mid \psi^\pm}}(\bzeta)= 
\chi_A^{\rho_{{\bf S}\mid \psi^\pm}}(\sqrt{\eta_R}\,\bzeta)e^{-(1-\eta_R)\bzeta^\dagger\bzeta},
\end{equation}
in the absence of background light, and it is 
\begin{equation}
\chi_A^{\rho_{\tilde{\bf S}\mid \psi^\pm}}(\bzeta) = 
\chi_A^{\rho_{{\bf S}\mid \psi^\pm}}(\sqrt{\eta_R}\,\bzeta)e^{-(1-\eta_R)(N_B +1)\bzeta^\dagger\bzeta}
\label{ZALMns''}
\end{equation}
in the presence of background light.  But now we must include the $M_B$ per polarization extraneous background modes that Alice and Bob's QRXs collect.  To do so we multiply
Eq.~(\ref{ZALMns''}) by 
\begin{equation}
\chi_A^{\rho_{\check{\mathcal{S}}}}(\bzeta_{\check{\mathcal{S}}}) = e^{-\bzeta_{\check{\mathcal{S}}}^\dagger\bzeta_{\check{\mathcal{S}}} (N_B+1)},
\end{equation}
where $\check{\mathcal{S}} \equiv \check{\mathcal{S}}_A\check{\mathcal{S}}_B$, with $\check{\mathcal{S}}_K \equiv \check{\mathcal{S}}_{K_1}\check{\mathcal{S}}_{K_2}\cdots\check{\mathcal{S}}_{K_{M_B}}$, for $K=A,B$, and $\check{\mathcal{S}}_{K_n} \equiv \check{\mathcal{S}}_{K_{n_H}}\check{\mathcal{S}}_{K_{n_V}}$ being the extraneous  background modes at Alice ($K=A$) and Bob's ($K=B$) QRXs.  
Equation~(\ref{loadableZALMnb}) follows readily from 
\begin{align}
\Pr({\rm loadable}) &= 1 - {\rm Tr}_{S_B\check{\mathcal{S}}_B}[{}_{S_A\check{\mathcal{S}}_A}\langle {\bf 0}|\hat{\rho}_{\tilde{\bf S}\mid \psi^-}\hat{\rho}_{\check{\mathcal{S}}}|{\bf 0}\rangle_{S_A\check{\mathcal{S}}_A}] 
- {\rm Tr}_{S_A\check{\mathcal{S}}_A}[{}_{S_B\check{\mathcal{S}}_B}\langle {\bf 0}|\hat{\rho}_{\tilde{\bf S}\mid \psi^-}\hat{\rho}_{\check{\mathcal{S}}}|{\bf 0}\rangle_{S_B\check{\mathcal{S}}_B}]  \nonumber \\[.05in]
& \,\,+ {}_{\tilde{\bf S}\check{\mathcal{S}}}\langle {\bf 0}|\hat{\rho}_{\tilde{\bf S}\mid \psi^-}\hat{\rho}_{\check{\mathcal{S}}}|{\bf 0}\rangle_{\tilde{\bf S}\check{\mathcal{S}}}
\end{align}
by doing the vacuum-state projections on the operator-valued inverse Fourier transform of $\chi_A^{\rho_{\tilde{\bf S}\mid \psi^\pm}}(\bzeta)\chi_A^{\rho_{\check{\mathcal{S}}}}(\bzeta_{\check{\mathcal{S}}})$ and taking the traces in the coherent-state basis.

To get ZALM's Bell-state probabilities in the presence of background, we note that 
$\hat{\rho}_{{\bf S},\check{\mathcal{S}}\mid I'_{\pm H}I'_{\mp V}}$ is the operator-valued inverse Fourier transform of
\begin{align}
\chi_A&^{\rho_{{\bf S},\check{\mathcal{S}}\mid I'_{\pm H}I'_{\mp V}}}(\bzeta,\bzeta_{\check{\mathcal{S}}}) = \nonumber \\[.05in]
&\,\,
e^{-\bzeta^\dagger \bzeta/N''_S}\left[1 - \frac{\eta_R|\zeta_{A_H}\pm\zeta_{B_H}|^2}{2N_S}\right] 
 \left[1 - \frac{\eta_R|\zeta_{A_V}\mp\zeta_{B_V}|^2}{2N_S}\right]e^{-\bzeta_{\check{\mathcal{S}}}^\dagger\bzeta_{\check{\mathcal{S}}} (N_B+1)}, 
\label{backgroundZALM}   
\end{align}
and that $\Pr(|\psi^-\rangle_{\tilde{\bf S}}\mid \psi^-)$ is the projection of that density operator onto $|\psi^-\rangle_{\tilde{\bf S}}|{\bf 0}\rangle_{\check{\mathcal{S}}}$.  Comparing Eq.~(\ref{backgroundZALM}) with Eq.~(\ref{prebackgroundZALM}) reveals that $\Pr(|\psi^-\rangle_{\tilde{\bf S}}\mid \psi^-)$ in the presence of background equals its no-background formula with $N_S'$ replaced by $N_S''$ and the result multiplied by $[1/(N_B+1)]^{4M_B}$.  The same procedure applies to ZALM's other Bell-state probabilities.  

Turning now to the Chahine~\emph{et al}.\@ source, we have that its anti-normally ordered characteristic function for the light collected by Alice and Bob's QRXs in the presence of background light is
\begin{align}
\chi^{\rho_{\tilde{\bf S}\check{\mathcal{S}} \mid \psi^-}}_A(\bzeta,\bzeta_{\check{\mathcal{S}}}) &= 
e^{-(|\zeta_{S_{A_H}}+\zeta_{S_{B_H}}|^2+|\zeta_{S_{A_V}}+\zeta_{S_{B_V}}|^2)/2N_S''}
e^{-(|\zeta_{S_{B_H}}-\zeta_{S_{B_H}}|^2+|\zeta_{S_{A_V}}-\zeta_{S_{B_V}}|^2)/2\tilde{N}_B} \nonumber \\[.05in]
& \,\,\times \left[1-\frac{\eta_R|\zeta_{S_{A_H}}+\zeta_{S_{B_H}}|^2}{2N_S}\right]\left[1-\frac{\eta_R|\zeta_{S_{A_V}}+\zeta_{S_{B_V}}|^2}{2N_S}\right]e^{-\bzeta_{\check{\mathcal{S}}}^\dagger\bzeta_{\check{\mathcal{S}}} (N_B+1)}.
\label{ChahineABnb}
\end{align}
We then get $\Pr({\rm loadable})$ from Eq.~(\ref{loadableChahineNb}) by following the steps just employed for ZALM with background light.  

Next, to find Chahine~\emph{et al}.'s Bell-state probabilities, we first recognize that the Bell-state probabilities equal Bell-state projections of the density operator for Alice and Bob's signal modes multiplied by vacuum-state projections of the extraneous background modes' density operator.  The latter supplies a factor of $[1/(N_B+1)]^{4M_B}$ and the former can be found using Appendix~\ref{AppendA}'s beam-splitter commutation technique starting from
\begin{equation}
\hat{\rho}_{S'S^{\prime\perp}\mid\psi^-} \equiv \int\!\frac{{\rm d}^8\bzeta}{\pi^4}\,
e^{-\bzeta_{S'}^\dagger\bzeta_{S'}/N_S''}e^{-\bzeta^\dagger_{S^{\prime\perp}}\bzeta_{S^{\prime\perp}}/N_B'}\left[1-\frac{\eta_R|\zeta_{S'_H}|^2}{N_S}\right]\left[1-\frac{\eta_R|\zeta_{S'_V}|^2}{N_S}\right]e^{-\bhata^\dagger\bzeta}e^{\bzeta^\dagger\bhata}.
\label{backpropagatedChahineNb}
\end{equation}
Doing the signal-modes' projection and multiplying by the extraneous background modes' term we arrive at Eqs.~(\ref{PrCorrectChahineNb})--(\ref{PrErrorChahineNb}). 

Finally, for unheralded operation in the presence of background, we note that Eq.~(\ref{chiAunheraldedRI}) applies with Eqs.~(\ref{tildeLamRnoNb}) and (\ref{tildeLamInoNb}) replaced by
\begin{equation}
\tilde{\Lambda}^{(R)}_{S_AI_B} = \left[\begin{array}{cc} \tilde{\Lambda}_{AA}  & \tilde{\Lambda}_{AB} \\[.05in]
\tilde{\Lambda}_{AB}^T & \tilde\Lambda_{BB}\end{array}\right],
\label{tildeLamRNb}
\end{equation}
and
\begin{equation}
\tilde{\Lambda}^{(I)}_{S_AI_B} = \left[\begin{array}{cc} \tilde{\Lambda}_{AA}  &  -\tilde{\Lambda}_{AB} \\[.05in]
-\tilde{\Lambda}_{AB}^T & \tilde\Lambda_{BB}\end{array}\right],
\label{tildeLamINb}
\end{equation}
respectively, where
\begin{equation}
\tilde{\Lambda}_{AA} = \tilde{\Lambda}_{BB} = \left[
\begin{array}{cc} 2[\eta_RG+(1-\eta_R)/N_B'] & 0  \\[.05in]
0 & 2[\eta_RG+(1-\eta_R)/N_B'] 
\end{array}\right],
\end{equation}
and
\begin{equation}
\tilde{\Lambda}_{AB} = \left[\begin{array}{cc}   0 & 2\eta_R\sqrt{G(G-1)} \\[.05in]
-2\eta_R\sqrt{G(G-1)} & 0 \end{array}\right].
\end{equation}
Having made those replacements we multiply the resulting $\chi_A^{\rho_{S_AI_B}}(\bzeta)$ by $e^{-\bzeta_{\check{\mathcal{S}}}^\dagger\bzeta_{\check{\mathcal{S}}}(N_B+1)}$, to get $\chi_A^{\rho_{S_AI_B\check{\mathcal{S}}}}(\bzeta,\bzeta_{\check{\mathcal{S}}})$.   

At this point we parallel what was done in Appendix~\ref{AppendB} to obtain  Eq.~(\ref{loadableUnheraldedNb}) from the operator-valued inverse Fourier transform of $\chi_A^{\rho_{S_AI_B\check{\mathcal{S}}}}(\bzeta,\bzeta_{\check{\mathcal{S}}})$ and 
\begin{align}
\Pr({\rm loadable}) &= 1 - {\rm Tr}_{I_B\check{\mathcal{S}}_B}[{}_{S_A\check{\mathcal{S}}_A}\langle {\bf 0}|\hat{\rho}_{S_AI_B\check{\mathcal{S}}}|{\bf 0}\rangle_{S_A\check{\mathcal{S}}_A}] - {\rm Tr}_{S_A\check{\mathcal{S}}_A}[{}_{I_B\check{\mathcal{S}}_B}\langle {\bf 0}|\hat{\rho}_{S_AI_B\check{\mathcal{S}}}|{\bf 0}\rangle_{I_B\check{\mathcal{S}}_B}]\nonumber \\[.05in]
&+ {}_{S_AI_B\check{\mathcal{S}}}\langle {\bf 0}|\hat{\rho}_{S_AI_B\check{\mathcal{S}}}|{\bf 0}\rangle_{S_AI_B\check{\mathcal{S}}}.
\end{align}
For unheralded operation's Bell-state probabilities in the presence of background we project the operator-valued inverse Fourier transform of $\chi_A^{\rho_{S_AI_B\check{\mathcal{S}}}}(\bzeta,\bzeta_{\check{\mathcal{S}}})$ onto
$|\psi^\pm\rangle_{S_AI_B}|{\bf 0}\rangle_{\check{\mathcal{S}}}$ and 
$|\phi^\pm\rangle_{S_AI_B}|{\bf 0}\rangle_{\check{\mathcal{S}}}$.  The extraneous background modes' term contributes a factor of $[1/(N_B+1)]^{4N_B}$ to each Bell-state probability.  The $S_AS_B$ modes' terms can be calculated in the same manner that was used in Appendix~\ref{AppendB} starting from Eq.~(\ref{chiAunheraldedRI}), with Eqs.~(\ref{tildeLamRnoNb}) and (\ref{tildeLamInoNb}) replaced by Eqs.~(\ref{tildeLamRNb}) and (\ref{tildeLamINb}).  The results so obtained complete the derivations of Eqs.~(\ref{PrCorrectUnheraldedNb}) and (\ref{PrErrorUnheraldedNb}).     

\acknowledgments
This work was supported by the Engineering Research
Centers Program of the National Science Foundation under
Grant No.\@ 1941583 to the NSF-ERC Center for Quantum Networks, and a NASA Space Technology Graduate Research Opportunity.   The authors thank Michael Raymer, Brian Smith, Franco Wong, and Saikat Guha for valuable discussions and encouragement, and Yousef Chahine for discussion of his signal-path erasure source.

\section*{Data Availability}
Data underlying the results presented in this paper are available in Ref.~\cite{data}.



\begin{thebibliography}{99}
\bibitem{footnote3}The partial-BSM apparatus only declares a herald when it detects 1 horizontally-polarized idler photon and 1 vertically-polarized idler photon.  With same-channel heralding, the detections must occur in the same wavelength channel.  With SPCI heralding, they can originate from any combination of the $N_I$ islands.
\bibitem{footnoteA}When the islands' collective bandwidth spans less than a few THz, phase-locked pumps for SPCI heralding's quantum-state frequency conversion  can be accomplished by filtering and amplifying a single ultrashort pump pulse.  For broader spans, however, two pump lasers will be needed, and their phase locking presents a significant additional implementation burden.
\bibitem{footnote6} An alternative expression for $\mathbb{E}[\mathcal{H}(N_M)]$, which follows from the easily obtained difference equation,
\begin{displaymath}
\mathbb{E}[\mathcal{H}(N_M+1)]-\mathbb{E}[\mathcal{H}_{N_M}] = 1- \sum_{k=0}^{N_M}p_k,\mbox{ with $\mathbb{E}[\mathcal{H}(1)] = 1-p_0$},
\end{displaymath} 
is
\begin{displaymath}
\mathbb{E}[\mathcal{H}(N_M)] = \sum_{k=0}^{N_M-1}\left(1-\sum_{\ell =0}^k p_\ell\right).
\end{displaymath}
\bibitem{footnote8}Here, and in all that follows, integrals without limits are from $-\infty$ to $\infty$ in all their dimensions.
\end{thebibliography}
\end{document}